\documentclass[aps,prb,showpacs,twocolumn,preprintnumbers,amsmath,amssymb]{revtex4}
\usepackage{dcolumn}
\usepackage{bm}
\usepackage{amsmath}
\usepackage{feynmp}

\usepackage{graphicx}

\begin{document}

\title{Quantum Interference Effects in Topological Nanowires In a Longitudinal Magnetic Field }

\author{Vincent E. Sacksteder IV$^{1}$}
\affiliation{W155 Wilson Building, Royal Holloway University of London, Egham Hill, Egham, TW20 0EX, United Kingdom }

\author{Quansheng Wu}
\affiliation{Theoretical Physics and Station Q Zurich, ETH Zurich, 8093 Zurich, Switzerland}
\email{wuquansheng@gmail.com }



\date{\today}

\begin{abstract}
We study  the  magnetoconductance of topological insulator nanowires in a longitudinal magnetic field, including Aharonov-Bohm, Altshuler-Aronov-Spivak, perfectly conducting channel, and universal conductance fluctuation effects. Our focus is on predicting experimental behavior in single wires in the quantum limit where temperature is reduced to zero. We show that changing the Fermi energy $E_F$  can tune a wire from from ballistic to diffusive conduction and to localization.   In both ballistic and diffusive single wires we find both Aharonov-Bohm and Altshuler-Aronov-Spivak oscillations with similar strengths, accompanied by quite strong universal conductance fluctuations (UCFs), all with amplitudes between $0.3 \, G_0$ and $1\,G_0$.  This contrasts strongly with the average behavior of many wires, which shows Aharonov-Bohm oscillations in the  ballistic regime and  Altshuler-Aronov-Spivak oscillations in the diffusive regime, with both oscillations substantially larger than the conductance fluctuations. In single wires the ballistic and diffusive regimes can be distinguished  by varying $E_F$ and studying the sign of the AB signal, which  depends periodically on $E_F$ in ballistic wires and randomly on $E_F$ in diffusive wires.  We also show that in long wires the perfectly conducting channel is visible at a wide range of energies within the bulk gap. We present typical conductance profiles at several wire lengths, showing that conductance fluctuations can dominate the average signal.  Similar behavior will be found in carbon nanotubes.  
\end{abstract}

\maketitle

 Strong topological insulators (TIs)  possess a band gap that can be used to eliminate electrical conduction through their interior, but unlike standard insulators they robustly host conducting surface states  which completely wrap all of the TI sample's surfaces.  \cite{Zhang09,RevModPhys.82.3045} This unique circumstance allows realization of the celebrated Aharonov-Bohm effect, where electrons are sensitive to the total magnetic flux through a specific loop.   \cite{PhysRev.115.485,RevModPhys.59.755}   If a TI wire has strictly constant cross-section along the wire's length, and the surface state has strictly zero penetration into the interior,  then the wire's conductance $G$ will be a strictly periodic function of the magnetic flux $\Phi$ through the wire's cross-section, i.e.  $G(\Phi) = G(\Phi + \Phi_0)$, where $\Phi_0 = h/e$ is the magnetic flux quantum.
  This periodic dependence on the total magnetic flux threading the electron's path, and not on any local details of the path, is the hallmark of the Aharonov-Bohm (AB) effect.

 A long string of experiments has realized the AB effect in TI wires, and has observed a  zoo of periodic conductance features.  One may distinguish between Aharonov-Bohm oscillations with period $ \Phi_0$ and   Altshuler-Aronov-Spivak (AAS) oscillations with period $\Phi_0/2$.  \cite{peng2010aharonov,xiu2011manipulating,lee2012single,li2012experimental,PhysRevLett.110.186806,sulaev2013experimental,Hamdou2013,tian2013dual,hamdou2013surface,safdar2013topological,zhu2014topological, zhu2014emergence,hong2014one,dufouleur2015pseudo,jauregui2016magnetic,jauregui2015gate,cho2015aharonov,kim2016quantum,kim2016quantumbeta}   In addition  Universal Conductance Fluctuations (UCFs) are observed - a noise-like   component of $G(\Phi)$ which depends sensitively on the Fermi level and on disorder. \cite{peng2010aharonov,PhysRevLett.106.196801, PhysRevB.85.075440,PhysRevB.88.155438,zhao2015electronic,PhysRevB.87.085442,li2014indications,kim2014weak,PhysRevLett.110.186806}  TIs also  host a Perfectly Conducting Channel (PCC) - a conductance quantum which is remarkable for its persistence in very long TI wires, its topological protection, and its status as a 3-D analogue of the quantum Hall effect.   \cite{PhysRevLett.69.1584,Mirlin94,Ando02,Takane04,PhysRevLett.99.116601,PhysRevLett.99.036601,ashitani2012perfectly, hong2014one, Wu14, kolomeisky2014aharonov, shimomura2015dephasing,PhysRevLett.105.206601, PhysRevLett.105.136403} 
 These effects are sensitive to  the scattering length $l$, the localization length $L_{LOC}$, the wire dimensions, Fermi level, and temperature. 
  This rather complex experimental situation is accompanied by a vast theoretical literature \cite{RevModPhys.57.287,RevModPhys.58.323,RevModPhys.69.731}
    on mesoscopic conduction which with a very  few exceptions \cite{PhysRevLett.105.206601,PhysRevLett.105.136403,PhysRevLett.105.156803,bardarson2013quantum,adroguer2012diffusion,kolomeisky2014aharonov}
predates topological insulators.

This paper offers an integrated view of these effects in the extreme quantum limit of zero temperature $T=0$, where the AB effect is most visible and has the most remarkable consequences.   This paper predicts the surface signal that experimentalists will see as they progressively implement improved TI devices with  stronger quantum interference.  If experiments eliminate bulk conduction, then only the surface signal described here will be observed; otherwise  an additional bulk signal will be observed.  We give special attention to the magnetoconductance's dependence on the Fermi level $E_F$ because it can be controlled systematically via gating. \cite{PhysRevLett.105.176602,wen2013electrostatic}  We also focus on single wires rather than the ensemble-averaged behavior of many wires, both because real experiments measure individual wires,
 and because ensemble averaging removes some of the most interesting aspects of the magnetoconductance.  This focus contrasts with previous works using ensemble averages which showed that period $\Phi_0$ AB oscillations are dominant in ballistic wires smaller than the scattering length $l$ and period $\Phi_0/2$ AAS oscillations are dominant in diffusive wires larger than $l$. \cite{PhysRevLett.105.156803,bardarson2013quantum}  In contrast, we show that single wires at $T=0$ manifest significant AB and AAS oscillations, as well as universal conduction fluctuations of the same or larger amplitude, regardless of whether they are ballistic or diffusive.

\textit{The Model. \label{TheModel}}
  We study TI nanowires fulfilling all the conditions necessary to ensure that the conductance be perfectly periodic throughout most of the bulk gap.   Briefly, these conditions are: a uniform cross-section,  neglegible penetration of the surface state into the bulk, a perfectly parallel magnetic field,  and the absence of bulk conduction. Any non-periodic component of the conductance implies violation of one of these conditions.    In particular, violation of the first three conditions generically  has the same result: the periodic signal remains unchanged at small magnetic fluxes, but is extinguished once the flux exceeds a threshold $N \Phi_0$.  The coefficient $N$ is infinite in an ideal TI wire and decreases as the wire quality worsens.  In appendix \ref{NonIdeal} we give simple estimates of $N$.  Based on these estimates, we expect that as long as the penetration depth and wire non-uniformity are less than a tenth of the wire radius, and the magnetic field is aligned with the wire axis within a tenth of a radian, $N \geq 5$ and the conductance should exhibit at least five $\Phi_0$ periods on either side of $B=0$ before being extinguished.  In particular, in the Bi$_2$Se$_3$ family of TIs which display a penetration depth $\lambda = 2-3$ nm, the wire radius should be of order $20$ nm.  \cite{zhang2010first} This is in fact the length scale chosen in many previous experiments on TI wires. \cite{peng2010aharonov}
  
 Violation of the last condition, the absence of bulk conduction, has much different effects.  Generically, bulk conduction will cause additional features in the conductance which are not periodic in the flux with period $\Phi_0$, so that the periodic surface signal of interest can be studied only after filtering out the nonperiodic signals.  The most notable such feature is an additional weak antilocalization conductance peak centered at zero flux, which has been reported in most experimental measurements of AB oscillations in TI wires.  In addition,  in long wires  bulk conduction will cause the topological surface state to tunnel through the bulk and be destroyed. \cite{Wu14} Two other frequently observed signatures of bulk conduction are a noise-like dependence on $\Phi$ which is slower  than $\Phi_0$, and an overall parabolic trend seen over many multiples of $\Phi_0$. Further discussion of these effects is outside the scope of this paper.

 We use  a  computationally efficient  minimal tight binding model of a strong  $\mathbb{Z}_2$  TI implemented on a cubic lattice.
   With four orbitals per site, the model's momentum representation is:
   \begin{align} \label{eqn:H}
      H = &  \sum_{i=1}^3 \left[\left(\imath  \frac{t}{2} \alpha_{i} 
                                         -\frac{1}{2}   \beta\right)e^{-\imath k_i a} + \rm{H.c.}\right]   + (m+3 ) \beta  
   \end{align}
\noindent $\alpha_{i} = \sigma_x \otimes \sigma_i$ and $\beta = \sigma_z \otimes 1$ are gamma matrices in the Dirac representation, $t =2$ is the hopping strength, $m=-1$ is the mass parameter,  and  $a=1$ is the lattice spacing.
\cite{liu2010model,ryu2012disorder,kobayashi2013disordered, kobayashi2014density} 
The large bulk band gap   $E= [-1,1]$ ensures that the surface state's penetration depth is very small, of order $O(a)$.   Our wires have constant height and width $h=w=10$, which is large enough to ensure that the topological state does not decay via tunneling through the bulk of the wire.  \cite{Wu14} 
 
  \begin{figure}[]
\includegraphics[width=8.5cm,clip,angle=0]{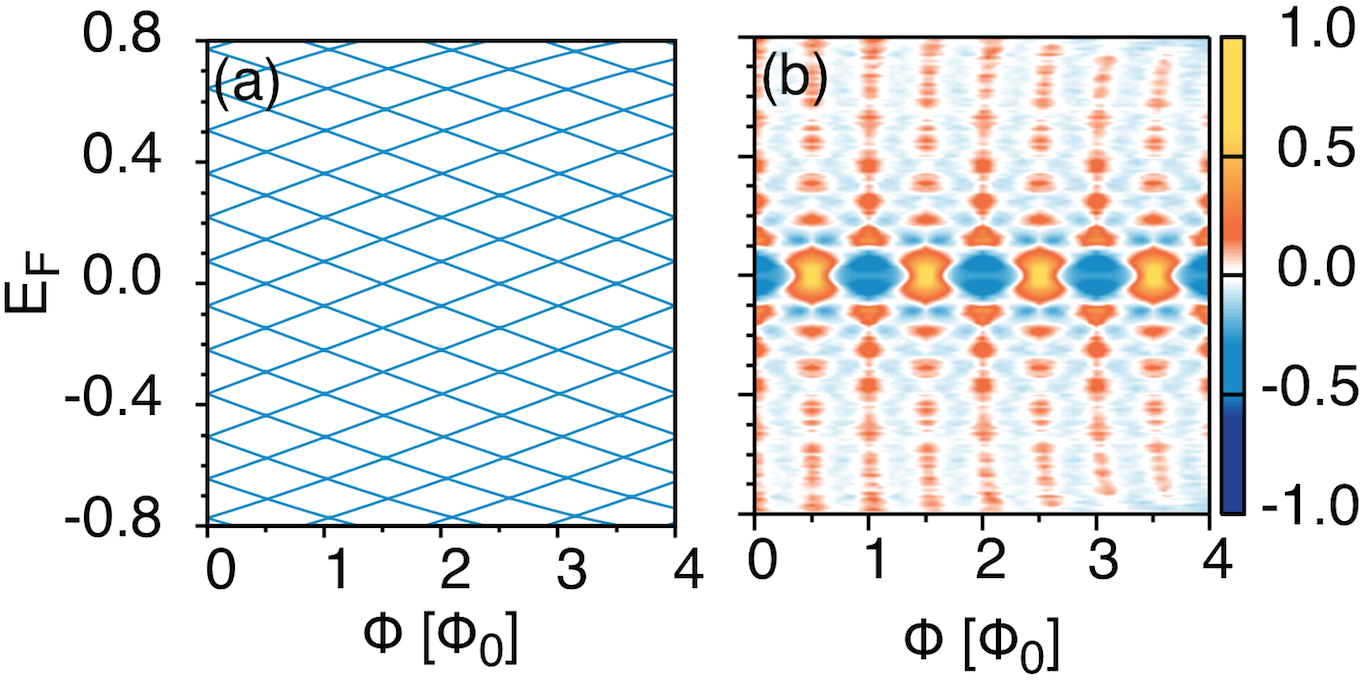}
\caption{ (Color online.)  The Aharonov-Bohm effect in $10 \times 10$ wires.  Pane (a) shows the surface state's spectrum in clean wires.     Pane (b) shows the ensemble average of the magnetoconductance $ G(E_F,\Phi)$ in $L_x = 67$ wires, with the average over $\Phi$ subtracted out.  The  regime of ballistic conduction is visible in the interval $E_{gap} < |E_F| < E_l$, where a cross-hatched pattern matches pane (a). At larger energies $E_l < |E_F|$ the vertical stripes show diffusive conduction.   Near the Dirac point $|E_F| <E_{gap}$ a small gap induced by  spin-momentum locking causes localization.    $E_{gap} =0.07$ is the maximal height of the gap reached at integer values of the magnetic flux, and $E_l = 0.35$ is the energy where the scattering length $l$ becomes smaller than the wire dimensions.    
}
\label{BandStructure}
\end{figure}

    To model scattering effects we add uncorrelated white noise  disorder $u(x)$  
chosen randomly from the interval $\left[ -W/2, \, W/2 \right] $, where $W=2$ is the disorder strength. The disorder is located only on the   TI sample's outer surface, and has a depth of one lattice unit. 
Our numerical studies have shown that as long as the Fermi level is not near the edges of bulk gap this  disorder's  qualitative and quantitative effects on the topological state are weak: the state remains tightly pinned to the surface and exhibits a fairly uniform surface density similar to that of a plane wave.  \cite{Wu14,PhysRevApplied.3.064006} Moreover its density of states, Fermi velocity, etc., do not undergo large changes. Our choice of weak surface disorder and no bulk disorder eliminates bulk conduction and minimizes the penetration depth  so that the AB effect is optimally realized. In real experiments with bulk disorder, as long as the Fermi level is not near the edges of the band gap, and as long as the disorder does not introduce carriers in the bulk of the wire,  the disorder's main effect will be a mild renormalization of the penetration depth.  It will not cause qualitative differences from the results presented here. 

 To calculate the conductance we use the Caroli formula $G = -G_0\;{Tr}((\Sigma^r_{L}-\Sigma^a_{L}) G^r_{LR} (\Sigma^r_{R}-\Sigma^a_{R}) G^a_{RL})$.  \cite{Caroli71,Meir92}   $G_0 = e^2/h$ is the conductance quantum, 
$\eta=10^{-9}$ regularizes the calculation,  $G^a, G^r = (E_F - H - u \mp \imath \epsilon)^{-1} $ are the advanced and retarded single-particle Green's functions connecting the left and right leads,  and $\Sigma_{L,R}$ are the lead self-energies.   For leads we minimize the contact resistance by using perfectly conducting 1-D wires connected to each site on the sample's ends, i.e.  at the two ends of the wire each orbital on each site is connected to a semi-infinite 1-D chain. The self-energy $\Gamma$ is simply $(2 t)^{-1} \exp(\imath \phi)$, where $t$ is the hopping strength within the leads and $\phi$ encodes the Fermi level.   \cite{Lopez84, lewenkopf2013recursive} 

 \textit{Control of conduction by changing the Fermi energy.}  In a 3-D TI the scattering length $l$ varies inversely with the Fermi energy.  Therefore  a single TI wire  can be ballistic, i.e. smaller than the scattering length, at  a small value of $E_F$, and at the same time diffusive, i.e. larger than the scattering length, at a larger $E_F$.  Figure \ref{BandStructure}b highlights  the  ballistic and the diffusive regimes in TI wires of length $L=67$.  It shows the ensemble-averaged magnetoconductance $G(E_F,\Phi) $, which is known to be dominated by period $\Phi_0$ AB oscillations in the  ballistic regime.  The \textbf{ballistic} regime is visible as a clear cross-hatched  pattern in the energy range $E_{gap} < |E_F| < E_l$, with $E_{gap} = 0.07, \, E_l = 0.35$. 
    This pattern is caused directly by the cross-hatched energy dispersion of the clean wire, shown in Figure \ref{BandStructure}a.   At $|E_F| = E_l$ the scattering length $l$ becomes smaller than the wire dimensions, and the wires become \textbf{diffusive}.   We show in appendix \ref{ScatteringLengthScaling} that $l \propto v_F^3   / W^2 \xi^2 E_F$, where $v_F = 2$ is the Fermi velocity, $W$ is the disorder strength, and $\xi$ is the disorder correlation length.  In individual wires the value of the energy $E_l$ separating the diffusive and ballistic regimes depends sensitively on the scattering length $l$, which is determined by the impurity type and concentration and requires experimental measurement on a case by case basis.  One way of determining $l$ is by placing leads at several distances along the wire length and measuring several resistances, and another is by measuring the Hall resistance. \cite{PhysRevB.88.041307}   In our wires $l$ is equal to the perimeter $P$ when $|E_F| = E_l = 0.35$.   Above this energy our wires are diffusive, so the ballistic cross-hatched pattern in Figure \ref{BandStructure}b is replaced by vertical stripes spaced at intervals of $\Phi_0/2$ - the well known AAS oscillations. 
    
 Figure \ref{BandStructure}b shows  much different physics  near the  Dirac point, at $|E_F| < E_{gap}$.  Here the conductance is never more than $G = 1\,G_0$, putting the sample in the \textbf{localized} regime  where quantum mechanical interference controls conduction.  When the flux has half-integer values $\Phi = (n+1/2)\Phi_0$ the Perfectly Conducting Channel is visible - a single conductance quantum $G = 1\,G_0$  which is topologically protected.  At other values of the flux the TI's  locking between spin and momentum, combined with the wire's finite size, opens a small gap around the Dirac point, causing the conductance to decrease exponentially with wire length.   The gap reaches its maximum height $E_{gap} = \sqrt{2} \,v_F/P \approx 0.07$ at integer flux $\Phi = n \, \Phi_0$,    where  $P=40$ is the wire's perimeter.  In a $50$ nm $\times \, 50$ nm Bi$_2$Se$_3$ wire, $E_{gap} \approx 14 $ meV, about one twentieth of the bulk band gap.

At energies $E_F$ outside   the  gap, i.e. $|E_F| > E_{gap}$ in our samples, the localization length $L_{LOC}\propto l \; |E_F|  / E_{gap} $. In individual wires $L_{LOC}$  will have to be determined on a case by case basis. Because $l$ varies inversely with the Fermi energy $E_F$, the localization length $L_{LOC} \propto v_F^2 \, P \,  / W^2 \xi^2$  is  independent of $E_F$.  (See appendix \ref{ScatteringLengthScaling} for a derivation.)  
 The only way to enter the localized regime, other than tuning  $E_F$ near the Dirac point, is to lengthen or narrow the wire, or increase the disorder.   

 \begin{figure}[]
\includegraphics[width=9.0cm,clip,angle=0]{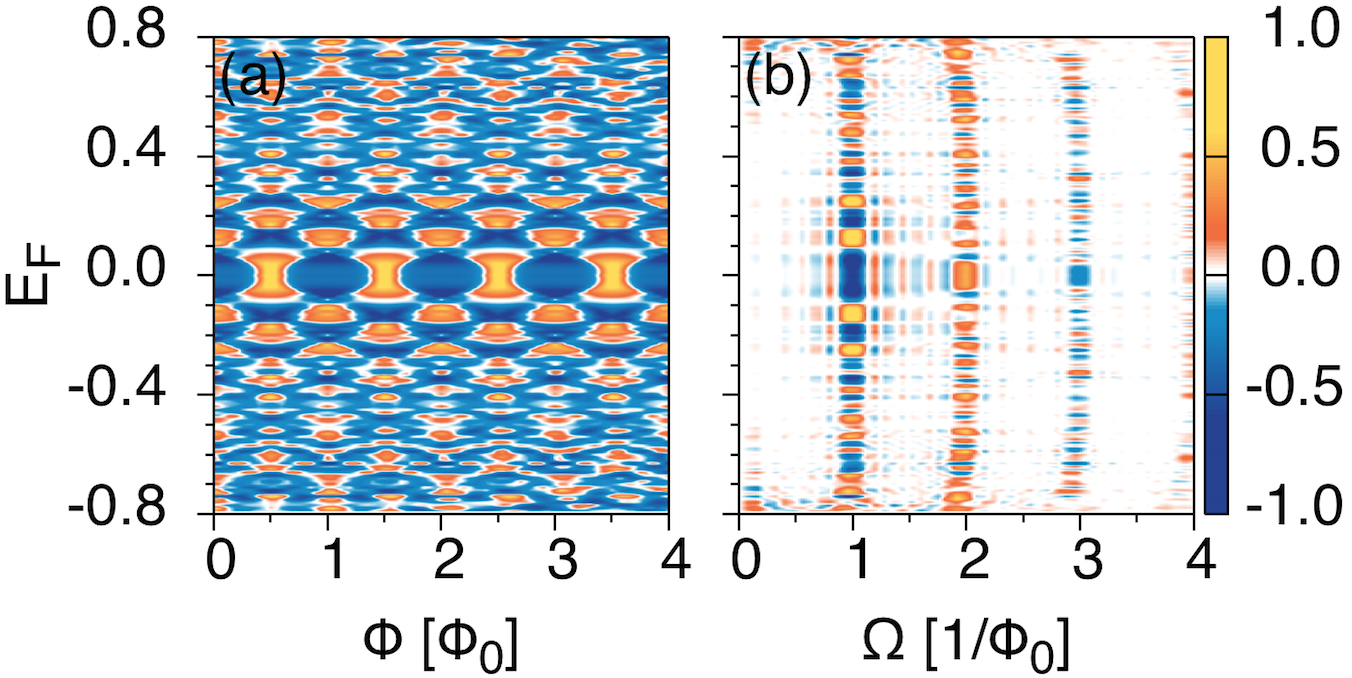}
\caption{ (Color online.) A single $L=67$ wire showing ballistic conduction at $E_{gap} < |E_F| < E_l$, diffusive conduction at $E_l < |E_F|$, and localization at $|E_F| < E_{gap}$.    The left pane shows the magnetoconductance $G(E_F,\Phi)$, and the right pane shows  its Fourier transform $ G(E_F,\Omega)$. The average over $\Phi$  has been removed.  $E_{gap} =0.07$ is the height of the gap, and $E_l = 0.35$ is the energy where the scattering length $l$ becomes smaller than the wire dimensions.
}
\label{ABAASEvsB}
\end{figure}

\textit{Single Wires.} Figure \ref{ABAASEvsB}a presents the magnetoconductance $G(E_F, \Phi)$ of a single $L=67$ wire.  The single-wire data manifests the same localized, ballistic, and diffusive regimes that are so clear in the ensemble-averaged $L=67$ wires presented earlier.
 It is however much more rich and detailed than the ensemble average, so we present the Fourier transform $ G(E_F, \Omega)$ of the conductance in  Figure \ref{ABAASEvsB}b, which affords a more precise analysis.  The Fourier transform is peaked at integer frequencies $\Omega = 1 / \Phi_0, \, 2/\Phi_0, \, 3/\Phi_0, ....$  and zero elsewhere, resulting in the vertical lines seen in pane (b).   The vertical line at $\Omega =1/\Phi_0$ shows AB oscillations, while the line at $\Omega = 2/\Phi_0$ shows  AAS oscillations.

Figure \ref{ABAASEvsB}b shows that in single wires at $T=0$, as opposed to ensembles of wires, AB oscillations can not be taken as a sign of ballistic conduction, and AAS oscillations are not a sign of diffusive conduction.  We find AB and AAS signals both in the ballistic regime at $|E_F| < E_l$ and in the  diffusive regime at $|E_F| > E_l$.  In the ballistic regime the AB amplitude oscillates periodically  as a function of $E_F$ in the range $\left[ -\, G_0, + \,G_0\right]$, producing the cross-hatched  pattern in Fig. \ref{ABAASEvsB}a.    In the diffusive regime the AB signal depends randomly on $E_F$ and generally remains in the range $\left[-0.4, 0.4 \right]\,G_0$.   In both the ballistic and diffusive regimes the AAS signal amplitude is a random function of $E_F$, and generally less than $0.4\,G_0$.   We conclude that  in single wires at  $T=0$ the presence or absence of AB and AAS signals can not be used to determine whether conduction is ballistic or diffusive.    The only way to determine this is to systematically vary the Fermi level $E_F$ and  determine whether the AB amplitude is a periodic function of $E_F$ as in the ballistic regime or instead random as in the diffusive regime.
 
 Our finding of AB oscillations in the diffusive regime confirm and extend a recent experiment   which found an AB signal in quite long wires  as long as the perimeter $P<l$ is less than than the scattering length $l$. \cite{dufouleur2015pseudo}  We find that if $P<l$  the AB signal depends periodically on $E_F$, and if $P>l$ its amplitude is random.   Our observation of AAS oscillations  verifies  theoretical work showing that  they occur in the ballistic regime as a consequence of constructive quantum interference between time-reversed circuits around the wire. \cite{kawabata1996altshuler,PhysRevB.57.6282,PhysRevB.69.205403,PhysRevB.70.161302}

The Fourier transform in pane \ref{ABAASEvsB}b also shows that  the AAS signal is predominantly positive across all values of the Fermi energy.  In  pane \ref{ABAASEvsB}a this means that the pattern of vertical AAS stripes has its maxima at  half integer flux $\Phi =  0, \Phi_0/2,  \Phi_0, 3 \Phi_0/2, ...$ and minima at quarter flux.  These minima and maxima are interchanged, and the Fourier transform's AAS signal is negative,  in  materials displaying weak localization.   The positive AAS signal seen in TIs is a direct indicator of weak antilocalization. 
 
 Comparison of the single-wire data in Figure \ref{ABAASEvsB} to the averaged data in  Figure \ref{BandStructure}b  shows that Universal Conductance Fluctuations, noise-like deviations  of single wires from the ensemble average, pervade both  the ballistic and diffusive regimes.  They occur  at the AB and AAS frequencies, and also at higher frequencies.   In our $T=0$ single wires  the UCF magnitude is $0.2 - 0.35 \, G_0$, so UCFs account for much of the total dependence on  $\Phi$ in single wires, including the entire  AB signal seen in the diffusive regime.   It is particularly remarkable that we find UCFs  in the ballistic regime, where most (but not all) of the electrons transiting the length of the wire do not scatter.   A single scattered electron is enough to change the conductance by $1\,G_0$.

 \begin{figure}[]
\includegraphics[width=9.0cm,clip,angle=0]{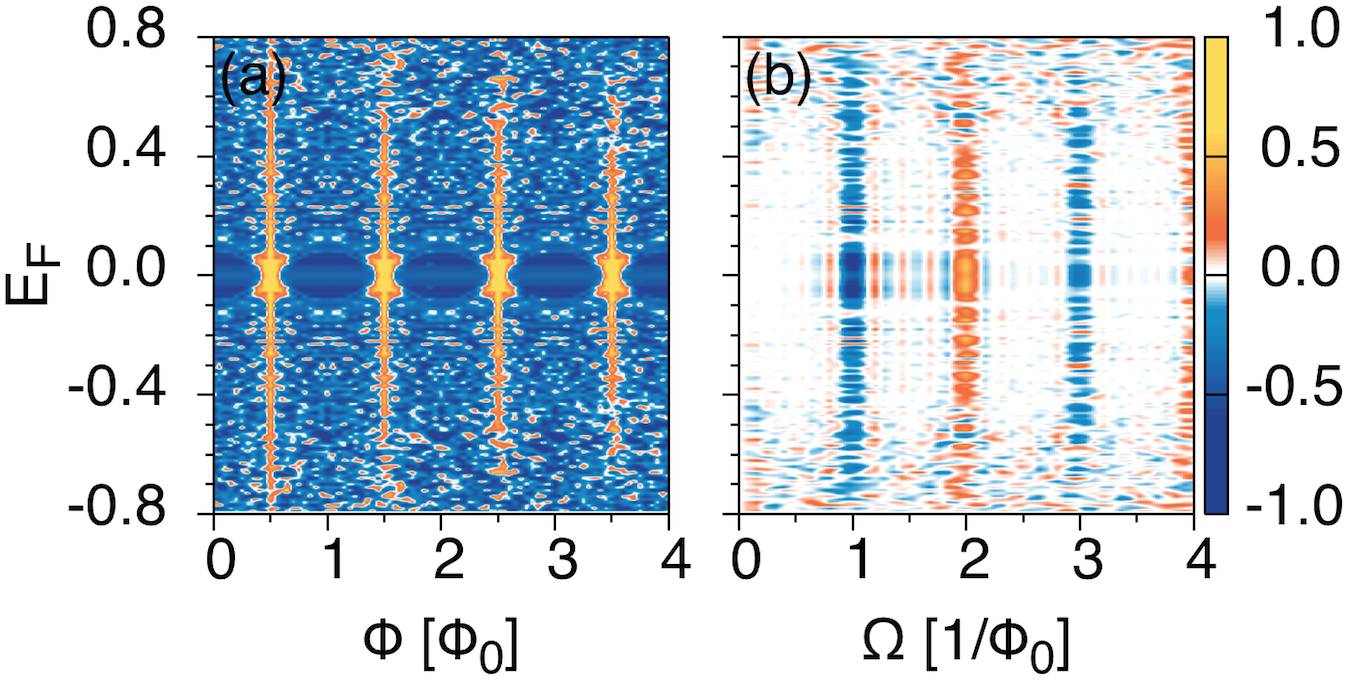}
\caption{ (Color online.)  A single long $L=403$ wire in the localized regime, showing the perfectly conducting channel (PCC) at $\Phi = \Phi_0/2, 3\Phi_0/2, 5\Phi_0/2, ...$.    The left pane shows the magnetoconductance $ G(E_F,\Phi)$, and the right pane shows   its Fourier transform $ G(E_F,\Omega)$. 
The average over $\Phi$ has been removed.  The localization length is $L_{LOC} \approx 200$.
}
\label{ABAASEvsBPCC}
\end{figure}

These results are expected to change with temperature.  Broadly speaking, the effect of temperature is to  reduce the total variation in $G(\Phi)$ and to remove frequencies in $G(\Omega)$, generally resulting in a smoother signal.  Although AB oscillations of order $1\,G_0$ have been observed in two experiments \cite{PhysRevLett.110.186806, cho2015aharonov}, 
 often a second scenario is observed where the measured signal is a small fraction of $G_0$ \cite{hong2014one,jauregui2016magnetic, kim2016quantum}, and very often only the AB signal is found.  Finite temperatures can cause this  scenario, regardless of whether the sample is ballistic or diffusive, by introducing a dephasing length $L_\phi$ beyond which quantum effects are extinguished by inelastic scattering. $L_\phi$ depends sensitively on both temperature and scattering, and in TIs has been measured to have values  from $100$ nm to microns.  \cite{PhysRevB.87.085442,PhysRevB.88.041307,PhysRevLett.110.186806}   If the dephasing length $L_\phi$ is smaller than the perimeter $P$ the magnetoconductance is exponentially suppressed, with the $N-th$ frequency $\Omega = N / \Phi_0$ controlled by $\exp(-N L_\phi / P)$.  \cite{RevModPhys.59.755,PhysRevLett.110.186806, hong2014one,kim2016quantum}  
A small value of $L_\phi/P < 1$ can explain experiments  where the total variation in $G(\Phi)$, in units of conductance, is substantially less than $1\,G_0$, and the AAS  signal is weak or absent.  In such experiments the AB signal should decrease exponentially with temperature, both in the ballistic and in the diffusive regime.  

If, the other hand, the inelastic dephasing length $L_\phi$ exceeds the wire perimeter, then the principal effect of temperature is via a second mechanism, smearing of the Fermi energy   $E_F$ over the thermal width $k_B T$.
  This thermal broadening can cause a substantial reduction in the AB signal, whose  sign is sensitive to $E_F$, while leaving the AAS signal relatively unscathed.
 In diffusive wires  with weak inelastic dephasing, thermal broadening  is expected to cause  the AB signal  to scale with $ 1/\sqrt{T}$, which has been confirmed by several experiments.   \cite{PhysRevB.32.4789,PhysRevB.64.045327,peng2010aharonov,xiu2011manipulating,lee2012single,Hamdou2013,sulaev2013experimental,jauregui2015gate}

\textit{Long Wires and the PCC.}  Figure \ref{ABAASEvsBPCC}  shows a single long wire in the localized regime, which  hosts a PCC.  The PCC  manifests as spectacular  very narrow vertical stripes with unit conductance extending through the  bulk gap and almost reaching the bulk band, so in long wires there is no need to tune the Fermi energy  to find the PCC.  The quantized conductance repeats at half-integer   flux $\Phi = (n+1/2)\Phi_0$.  At other values of $\Phi$ the conductance decays exponentially with wire length as is typical in localized wires, so the PCC stripes are very narrow.    This  ensures that many frequencies are present in the Fourier transform (Figure \ref{ABAASEvsBPCC}b), and that the even frequencies $\Omega = 0, 2/\Phi_0, 4/\Phi_0, ...$ have negative sign while the odd frequencies $\Omega = \Phi_0,3/\Phi_0,...$ have positive sign.  The peak sharpness, and also the number of frequencies in the Fourier transform, increases with the wire length. The quantization of the PCC peaks is controlled by their magnetic-flux-induced decay length, which far exceeds the localization length and scales with the cube of the wire width $w^3$. \cite{Wu14} 

As discussed earlier and shown in Figure \ref{ABAASEvsB}, the PCC can be observed  in short wires near the Dirac point.     However  in this case the PCC peaks are  roughly sinusoidal and match well with a basic AB signal. 
   
   To our knowledge, in 3-D TIs the PCC's sensitivity to  temperature has not yet been studied.  Graphene ribbons with zigzag edges exhibit a pair of PCCs, one for each of graphene's two valleys, resulting in a $2 G_0$ conductance.  These PCCs are known to be unstable against dephasing via a mechanism which mixes the two valleys \cite{Ando02, shimomura2015dephasing}.  We emphasize here that the PCC in 3-D TIs is not vulnerable to the same decay mechanism, because there is an odd number of conducting channels and a single PCC.  \cite{PhysRevLett.105.136403, PhysRevLett.105.206601} In particular, dephasing per se, i.e. randomization of the wave-function's phase, can affect only short wires where  the conductance is larger than $1\, G_0$.  In these samples such dephasing  will eliminate weak antilocalization, which in the absence of dephasing multiplies the conductance by $\ln L/l$.  Here  $L$ is the sample size and $l$ is the scattering length.  In longer wires where only the PCC remains and the conductance is quantized at $1 \,G_0$, dephasing per se cannot produce any further effect on the single remaining channel.  \cite{takane2010anomalous,ashitani2012perfectly, shimomura2015dephasing}  However, unlike a truly one-channel topological wire, the surfaces of quasi-one-dimensional 3-D TI wires do host localized states.  There is some possibility that inelastic many-body processes might be able to couple those states to the  PCC and eventually  destroy the PCC.   Further study of this possibility would require a careful perturbative treatment of interactions similar to the analysis applied to 2-D TI edge states in Refs. \onlinecite{PhysRevLett.108.156402,PhysRevB.90.075118}, combined with careful numerical analysis of both localized and PCC states in long TI wires.  Such analysis is outside the scope of the present article.
   
      \begin{figure}[]
\includegraphics[width=9.0cm,clip,angle=0]{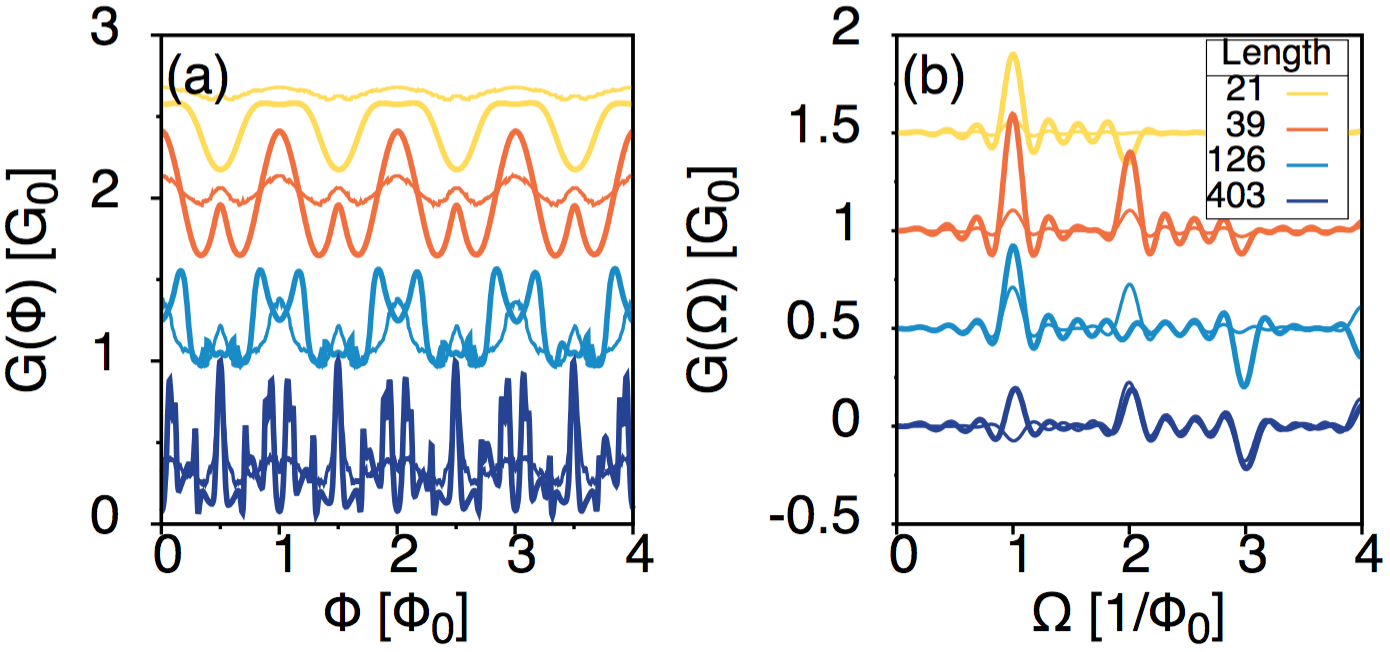}
\caption{ (Color online.) Magnetoconductance $G(\Phi)$  and its Fourier transform $G(\Omega)$ at four wire lengths corresponding to ultrashort, ballistic, diffusive, and localized wires.  Thick lines show the  conductance  of a single wire and thin lines show the ensemble average.    The FFT lines have been shifted to allow comparison and the $\Omega=0$ component has been removed.  $E_F = 0.26$ and the scattering and localization lengths are $l \approx 50, \;  L_{LOC} \approx 200$. 
}
\label{LinePlots}
\end{figure}
Figure \ref{LinePlots} summarizes typical magnetoconductance profiles at fixed $E_F$ and zero temperature in ultrashort (yellow), ballistic (red), diffusive (blue), and localized (purple) wires.  Single wire results are shown in bold, and the ensemble average is shown with thin lines.  The AB oscillations seen in the ensemble average of the diffusive (blue) wires are caused by the perimeter $P$ being less than the scattering length $l$, and will disappear in the opposite case of $P > l$.     At each of the four wire lengths the UCF strength, i.e. the standard deviation of $G(\Phi)$, is  $0.2 - 0.35 \, G_0$.  In ultrashort and ballistic wires the max vs. min of the ensemble average is $0.08 \, G_0$ and $0.18\,G_0$ respectively, which is small compared to the UCFs.  Therefore in single wires at fixed $E_F$ the amplitude of the AB and AAS signals has a very strong  random component.   In diffusive and localized wires the ensemble max vs. min  grows to $0.42\,G_0$ and $0.75\,G_0$, so that single wire results  get closer to the average behavior, albeit with still strong randomness. In particular single localized wires should reliably manifest a picket fence pattern of  PCC peaks.

Lastly we point out that the longitudinal  magnetoconductance of carbon nanotubes \cite{strunk2006quantum,stojetz2007competition}  is quite similar to that of  the TI wires studied here, with only three differences.
   Firstly, carbon nanotubes exhibit four species of Dirac particles, in two valleys and with two distinct spins, multiplying the ballistic conductance and the PCC by four.  Secondly, the  perfectly conducting channel occurs at integer flux, not half-integer flux; no magnetic field is required to see the PCC in carbon nanotubes.  Thirdly, if short range scattering or interactions mix the two graphene valleys, then the valley mixing will kill the PCC and reverse the sign of the AAS signal.
  Aside from these details, the  magnetoconductance should be similar to that of TI wires, including the relative strength of the  various effects.
   


\appendix


\section{The scattering and localization lengths \label{ScatteringLengthScaling}}
The scattering length $l$ can be obtained from the self-consistent Born approximation $\hbar / \tau = \gamma \rho$, where $\tau$ is the scattering time, $\gamma$ is the scattering strength, and $\rho$ is the density of states.    The topological surface state obeys  a 2-D Dirac dispersion, resulting in a density of states $\rho \propto E_F / v_F^2$ which is proportional to the Fermi energy $E_F$, where $v_F$ is the Fermi velocity.  The scattering time $\tau$ is related to the scattering length $l$ by $\tau = l / v_F$.  Using $\gamma \propto W^2  \xi^2$, where $W$ is the disorder strength and $\xi$ is the disorder correlation length, and setting $\hbar = 1$, we obtain the scattering length formula given in the text, $l \propto  v_F^3 / W^2 \xi^2  E_F$.

The localization length $L_{LOC}$ can be estimated from $L_{LOC} \propto l \; N$, where $N \propto |E_F| / E_{gap}$ is the number of conducting channels in a clean wire.  In a 2-D electron gas of width $w$ the number of conducting channels is $N \propto w \rho\, v_F$.  On the surface of a TI wire $w$ is equal to the wire perimeter $P$.   Using again $\rho \propto E_F / v_F^2$, we obtain the localization length formula given in the text, $L_{LOC} \propto  v_F^2 P / W^2 \xi^2$.

\section{Effects of Non-Ideal Nanowires on the AB effect \label{NonIdeal}}
We consider the effect of a non-uniform wire cross-section, penetration of the surface state into the bulk, and the presence of a magnetic field component perpendicular to the axis of the TI wire. Each of these three effects add an additional length scale $\lambda$ to the wire:
\begin{itemize}
\item Non-uniform cross-section: For each cross-section $A$ we can calculate an effective radius $r = \sqrt{A/\pi}$.  The new length scale $\lambda$ is the difference between $r$'s minimum and maximum values, i.e. $\lambda = r_{max} - r_{min}$.  
\item Penetration into the bulk: Here $\lambda$ is  the penetration depth. \cite{RevModPhys.59.755}
\item Perpendicular magnetic field: In this case $\lambda$ is  the radius $r$ of the wire multiplied by the sin of the angle of the total magnetic field with respect to the wire axis, i.e. $\lambda =r \, \sin \theta$.  \cite{RevModPhys.59.755}
\end{itemize}
In ideal wires $\lambda$ is zero.   Assuming that the wire is not too far from perfection, i.e.  $\lambda$ is small compared to the wire radius,  we  calculate the magnetic flux  $\Phi_\lambda$ through the portion of the wire which is affected by the wire imperfection.  The cross-section of this imperfect part is $2 \pi r \lambda$.   Assuming that the wire's total cross-section is $\pi r^2$, and that a total flux $\Phi$ passes through the cross-section of the wire, we find that $\Phi_\lambda = (2 \lambda /r ) \Phi$ flux units pass through the imperfect part of the wire.

It is the $\Phi_\lambda$ flux which is sensitive to the wire's imperfections, and which multiplies the signal by a random phase $\exp(\imath \Phi_\lambda / \Phi_0)$.  As long as the phase is small, i.e.   $\Phi_\lambda < \Phi_0$ is less than one flux quantum, the imperfection has little effect.  However once $\Phi_\lambda$ exceeds one flux quantum, the imperfection is able to completely randomize the phase and destroy the Aharonov-Bohm effect.  Therefore we identify the threshold value of the total magnetic flux as $N \Phi_0$, where $N = r / 2\lambda$, $r$ is the wire radius, and $\lambda$ is given above.  In particular, when the magnetic field is not perfectly parallel to the TI wire, $N = 1 / 2 \sin \theta$.  The first $N$ periodic oscillations of the conductance will be easily visible in the experimental data,  while higher oscillations will be extinguished.


 \begin{acknowledgments}  
 We gratefully acknowledge useful discussions with Xi Dai, Zhong Fang, ShengNan Zhang, Hongmeng Weng, Stefan Kettemann,  Tomi Ohtsuki, Igor Gornyi, and Alexander Mirlin.  This work was initiated at the Institute of Physics in Beijing with support from the National Science Foundation of China and the 973 program of China under Contract No. 2011CBA00108 and Contract No.11404024.  Q.S. Wu was supported by  Microsoft Research and the Swiss National Science Foundation through the National Competence Centers in Research MARVEL.
\end{acknowledgments}

\bibliography{Vincent}

\begin{thebibliography}{72}
\expandafter\ifx\csname natexlab\endcsname\relax\def\natexlab#1{#1}\fi
\expandafter\ifx\csname bibnamefont\endcsname\relax
  \def\bibnamefont#1{#1}\fi
\expandafter\ifx\csname bibfnamefont\endcsname\relax
  \def\bibfnamefont#1{#1}\fi
\expandafter\ifx\csname citenamefont\endcsname\relax
  \def\citenamefont#1{#1}\fi
\expandafter\ifx\csname url\endcsname\relax
  \def\url#1{\texttt{#1}}\fi
\expandafter\ifx\csname urlprefix\endcsname\relax\def\urlprefix{URL }\fi
\providecommand{\bibinfo}[2]{#2}
\providecommand{\eprint}[2][]{\url{#2}}

\bibitem[{\citenamefont{Zhang et~al.}(2009)\citenamefont{Zhang, Liu, Qi, Dai,
  Fang, and Zhang}}]{Zhang09}
\bibinfo{author}{\bibfnamefont{H.}~\bibnamefont{Zhang}},
  \bibinfo{author}{\bibfnamefont{C.-X.} \bibnamefont{Liu}},
  \bibinfo{author}{\bibfnamefont{X.-L.} \bibnamefont{Qi}},
  \bibinfo{author}{\bibfnamefont{X.}~\bibnamefont{Dai}},
  \bibinfo{author}{\bibfnamefont{Z.}~\bibnamefont{Fang}}, \bibnamefont{and}
  \bibinfo{author}{\bibfnamefont{S.-C.} \bibnamefont{Zhang}},
  \bibinfo{journal}{Nature Physics} \textbf{\bibinfo{volume}{5}},
  \bibinfo{pages}{438} (\bibinfo{year}{2009}).

\bibitem[{\citenamefont{Hasan and Kane}(2010)}]{RevModPhys.82.3045}
\bibinfo{author}{\bibfnamefont{M.~Z.} \bibnamefont{Hasan}} \bibnamefont{and}
  \bibinfo{author}{\bibfnamefont{C.~L.} \bibnamefont{Kane}},
  \bibinfo{journal}{Rev. Mod. Phys.} \textbf{\bibinfo{volume}{82}},
  \bibinfo{pages}{3045} (\bibinfo{year}{2010}).

\bibitem[{\citenamefont{Aharonov and Bohm}(1959)}]{PhysRev.115.485}
\bibinfo{author}{\bibfnamefont{Y.}~\bibnamefont{Aharonov}} \bibnamefont{and}
  \bibinfo{author}{\bibfnamefont{D.}~\bibnamefont{Bohm}},
  \bibinfo{journal}{Phys. Rev.} \textbf{\bibinfo{volume}{115}},
  \bibinfo{pages}{485} (\bibinfo{year}{1959}).

\bibitem[{\citenamefont{Aronov and Sharvin}(1987)}]{RevModPhys.59.755}
\bibinfo{author}{\bibfnamefont{A.~G.} \bibnamefont{Aronov}} \bibnamefont{and}
  \bibinfo{author}{\bibfnamefont{Y.~V.} \bibnamefont{Sharvin}},
  \bibinfo{journal}{Review of Modern Physics} \textbf{\bibinfo{volume}{59}},
  \bibinfo{pages}{755} (\bibinfo{year}{1987}).

\bibitem[{\citenamefont{Peng et~al.}(2010)\citenamefont{Peng, Lai, Kong,
  Meister, Chen, Qi, Zhang, Shen, and Cui}}]{peng2010aharonov}
\bibinfo{author}{\bibfnamefont{H.}~\bibnamefont{Peng}},
  \bibinfo{author}{\bibfnamefont{K.}~\bibnamefont{Lai}},
  \bibinfo{author}{\bibfnamefont{D.}~\bibnamefont{Kong}},
  \bibinfo{author}{\bibfnamefont{S.}~\bibnamefont{Meister}},
  \bibinfo{author}{\bibfnamefont{Y.}~\bibnamefont{Chen}},
  \bibinfo{author}{\bibfnamefont{X.-L.} \bibnamefont{Qi}},
  \bibinfo{author}{\bibfnamefont{S.-C.} \bibnamefont{Zhang}},
  \bibinfo{author}{\bibfnamefont{Z.-X.} \bibnamefont{Shen}}, \bibnamefont{and}
  \bibinfo{author}{\bibfnamefont{Y.}~\bibnamefont{Cui}},
  \bibinfo{journal}{Nature Mat.} \textbf{\bibinfo{volume}{9}},
  \bibinfo{pages}{225} (\bibinfo{year}{2010}).

\bibitem[{\citenamefont{Xiu et~al.}(2011)\citenamefont{Xiu, He, Wang, Cheng,
  Chang, Lang, Huang, Kou, Zhou, Jiang et~al.}}]{xiu2011manipulating}
\bibinfo{author}{\bibfnamefont{F.}~\bibnamefont{Xiu}},
  \bibinfo{author}{\bibfnamefont{L.}~\bibnamefont{He}},
  \bibinfo{author}{\bibfnamefont{Y.}~\bibnamefont{Wang}},
  \bibinfo{author}{\bibfnamefont{L.}~\bibnamefont{Cheng}},
  \bibinfo{author}{\bibfnamefont{L.-T.} \bibnamefont{Chang}},
  \bibinfo{author}{\bibfnamefont{M.}~\bibnamefont{Lang}},
  \bibinfo{author}{\bibfnamefont{G.}~\bibnamefont{Huang}},
  \bibinfo{author}{\bibfnamefont{X.}~\bibnamefont{Kou}},
  \bibinfo{author}{\bibfnamefont{Y.}~\bibnamefont{Zhou}},
  \bibinfo{author}{\bibfnamefont{X.}~\bibnamefont{Jiang}},
  \bibnamefont{et~al.}, \bibinfo{journal}{Nature Nanotech.}
  \textbf{\bibinfo{volume}{6}}, \bibinfo{pages}{216} (\bibinfo{year}{2011}).

\bibitem[{\citenamefont{Lee et~al.}(2012)\citenamefont{Lee, In, Yoo, Jo, Park,
  Kim, Koo, Kim, Kim, and Wang}}]{lee2012single}
\bibinfo{author}{\bibfnamefont{S.}~\bibnamefont{Lee}},
  \bibinfo{author}{\bibfnamefont{J.}~\bibnamefont{In}},
  \bibinfo{author}{\bibfnamefont{Y.}~\bibnamefont{Yoo}},
  \bibinfo{author}{\bibfnamefont{Y.}~\bibnamefont{Jo}},
  \bibinfo{author}{\bibfnamefont{Y.~C.} \bibnamefont{Park}},
  \bibinfo{author}{\bibfnamefont{H.-J.} \bibnamefont{Kim}},
  \bibinfo{author}{\bibfnamefont{H.~C.} \bibnamefont{Koo}},
  \bibinfo{author}{\bibfnamefont{J.}~\bibnamefont{Kim}},
  \bibinfo{author}{\bibfnamefont{B.}~\bibnamefont{Kim}}, \bibnamefont{and}
  \bibinfo{author}{\bibfnamefont{K.~L.} \bibnamefont{Wang}},
  \bibinfo{journal}{Nano Lett.} \textbf{\bibinfo{volume}{12}},
  \bibinfo{pages}{4194} (\bibinfo{year}{2012}).

\bibitem[{\citenamefont{Li et~al.}(2012)\citenamefont{Li, Qin, Song, Wang,
  Wang, Wang, Ding, Van~Haesondonck, Wan, Zhang et~al.}}]{li2012experimental}
\bibinfo{author}{\bibfnamefont{Z.}~\bibnamefont{Li}},
  \bibinfo{author}{\bibfnamefont{Y.}~\bibnamefont{Qin}},
  \bibinfo{author}{\bibfnamefont{F.}~\bibnamefont{Song}},
  \bibinfo{author}{\bibfnamefont{Q.-H.} \bibnamefont{Wang}},
  \bibinfo{author}{\bibfnamefont{X.}~\bibnamefont{Wang}},
  \bibinfo{author}{\bibfnamefont{B.}~\bibnamefont{Wang}},
  \bibinfo{author}{\bibfnamefont{H.}~\bibnamefont{Ding}},
  \bibinfo{author}{\bibfnamefont{C.}~\bibnamefont{Van~Haesondonck}},
  \bibinfo{author}{\bibfnamefont{J.}~\bibnamefont{Wan}},
  \bibinfo{author}{\bibfnamefont{Y.}~\bibnamefont{Zhang}},
  \bibnamefont{et~al.}, \bibinfo{journal}{App. Phys. Lett.}
  \textbf{\bibinfo{volume}{100}}, \bibinfo{pages}{083107}
  (\bibinfo{year}{2012}).

\bibitem[{\citenamefont{Dufouleur et~al.}(2013)\citenamefont{Dufouleur, Veyrat,
  Teichgr\"aber, Neuhaus, Nowka, Hampel, Cayssol, Schumann, Eichler, Schmidt
  et~al.}}]{PhysRevLett.110.186806}
\bibinfo{author}{\bibfnamefont{J.}~\bibnamefont{Dufouleur}},
  \bibinfo{author}{\bibfnamefont{L.}~\bibnamefont{Veyrat}},
  \bibinfo{author}{\bibfnamefont{A.}~\bibnamefont{Teichgr\"aber}},
  \bibinfo{author}{\bibfnamefont{S.}~\bibnamefont{Neuhaus}},
  \bibinfo{author}{\bibfnamefont{C.}~\bibnamefont{Nowka}},
  \bibinfo{author}{\bibfnamefont{S.}~\bibnamefont{Hampel}},
  \bibinfo{author}{\bibfnamefont{J.}~\bibnamefont{Cayssol}},
  \bibinfo{author}{\bibfnamefont{J.}~\bibnamefont{Schumann}},
  \bibinfo{author}{\bibfnamefont{B.}~\bibnamefont{Eichler}},
  \bibinfo{author}{\bibfnamefont{O.~G.} \bibnamefont{Schmidt}},
  \bibnamefont{et~al.}, \bibinfo{journal}{Phys. Rev. Lett.}
  \textbf{\bibinfo{volume}{110}}, \bibinfo{pages}{186806}
  (\bibinfo{year}{2013}).

\bibitem[{\citenamefont{Sulaev et~al.}(2013)\citenamefont{Sulaev, Ren, Xia,
  Lin, Yu, Qiu, Zhang, Han, Li, Zhu et~al.}}]{sulaev2013experimental}
\bibinfo{author}{\bibfnamefont{A.}~\bibnamefont{Sulaev}},
  \bibinfo{author}{\bibfnamefont{P.}~\bibnamefont{Ren}},
  \bibinfo{author}{\bibfnamefont{B.}~\bibnamefont{Xia}},
  \bibinfo{author}{\bibfnamefont{Q.~H.} \bibnamefont{Lin}},
  \bibinfo{author}{\bibfnamefont{T.}~\bibnamefont{Yu}},
  \bibinfo{author}{\bibfnamefont{C.}~\bibnamefont{Qiu}},
  \bibinfo{author}{\bibfnamefont{S.-Y.} \bibnamefont{Zhang}},
  \bibinfo{author}{\bibfnamefont{M.-Y.} \bibnamefont{Han}},
  \bibinfo{author}{\bibfnamefont{Z.~P.} \bibnamefont{Li}},
  \bibinfo{author}{\bibfnamefont{W.~G.} \bibnamefont{Zhu}},
  \bibnamefont{et~al.}, \bibinfo{journal}{AIP Adv.}
  \textbf{\bibinfo{volume}{3}}, \bibinfo{pages}{032123} (\bibinfo{year}{2013}).

\bibitem[{\citenamefont{Hamdou et~al.}(2013{\natexlab{a}})\citenamefont{Hamdou,
  Gooth, Dorn, Pippel, and Nielsch}}]{Hamdou2013}
\bibinfo{author}{\bibfnamefont{B.}~\bibnamefont{Hamdou}},
  \bibinfo{author}{\bibfnamefont{J.}~\bibnamefont{Gooth}},
  \bibinfo{author}{\bibfnamefont{A.}~\bibnamefont{Dorn}},
  \bibinfo{author}{\bibfnamefont{E.}~\bibnamefont{Pippel}}, \bibnamefont{and}
  \bibinfo{author}{\bibfnamefont{K.}~\bibnamefont{Nielsch}},
  \bibinfo{journal}{Appl. Phys. Lett.} \textbf{\bibinfo{volume}{102}},
  \bibinfo{pages}{223110} (\bibinfo{year}{2013}{\natexlab{a}}).

\bibitem[{\citenamefont{Tian et~al.}(2013)\citenamefont{Tian, Ning, Qu, Du,
  Wang, and Zhang}}]{tian2013dual}
\bibinfo{author}{\bibfnamefont{M.}~\bibnamefont{Tian}},
  \bibinfo{author}{\bibfnamefont{W.}~\bibnamefont{Ning}},
  \bibinfo{author}{\bibfnamefont{Z.}~\bibnamefont{Qu}},
  \bibinfo{author}{\bibfnamefont{H.}~\bibnamefont{Du}},
  \bibinfo{author}{\bibfnamefont{J.}~\bibnamefont{Wang}}, \bibnamefont{and}
  \bibinfo{author}{\bibfnamefont{Y.}~\bibnamefont{Zhang}},
  \bibinfo{journal}{Sci. Rep.} \textbf{\bibinfo{volume}{3}}
  (\bibinfo{year}{2013}).

\bibitem[{\citenamefont{Hamdou et~al.}(2013{\natexlab{b}})\citenamefont{Hamdou,
  Gooth, Dorn, Pippel, and Nielsch}}]{hamdou2013surface}
\bibinfo{author}{\bibfnamefont{B.}~\bibnamefont{Hamdou}},
  \bibinfo{author}{\bibfnamefont{J.}~\bibnamefont{Gooth}},
  \bibinfo{author}{\bibfnamefont{A.}~\bibnamefont{Dorn}},
  \bibinfo{author}{\bibfnamefont{E.}~\bibnamefont{Pippel}}, \bibnamefont{and}
  \bibinfo{author}{\bibfnamefont{K.}~\bibnamefont{Nielsch}},
  \bibinfo{journal}{Appl. Phys. Lett.} \textbf{\bibinfo{volume}{103}},
  \bibinfo{pages}{193107} (\bibinfo{year}{2013}{\natexlab{b}}).

\bibitem[{\citenamefont{Safdar et~al.}(2013)\citenamefont{Safdar, Wang, Mirza,
  Wang, Xu, and He}}]{safdar2013topological}
\bibinfo{author}{\bibfnamefont{M.}~\bibnamefont{Safdar}},
  \bibinfo{author}{\bibfnamefont{Q.}~\bibnamefont{Wang}},
  \bibinfo{author}{\bibfnamefont{M.}~\bibnamefont{Mirza}},
  \bibinfo{author}{\bibfnamefont{Z.}~\bibnamefont{Wang}},
  \bibinfo{author}{\bibfnamefont{K.}~\bibnamefont{Xu}}, \bibnamefont{and}
  \bibinfo{author}{\bibfnamefont{J.}~\bibnamefont{He}}, \bibinfo{journal}{Nano
  Lett.} \textbf{\bibinfo{volume}{13}}, \bibinfo{pages}{5344}
  (\bibinfo{year}{2013}).

\bibitem[{\citenamefont{Zhu et~al.}(2014{\natexlab{a}})\citenamefont{Zhu, Zhao,
  Richter, and Li}}]{zhu2014topological}
\bibinfo{author}{\bibfnamefont{H.}~\bibnamefont{Zhu}},
  \bibinfo{author}{\bibfnamefont{E.}~\bibnamefont{Zhao}},
  \bibinfo{author}{\bibfnamefont{C.~A.} \bibnamefont{Richter}},
  \bibnamefont{and} \bibinfo{author}{\bibfnamefont{Q.}~\bibnamefont{Li}},
  \bibinfo{journal}{ECS Trans.} \textbf{\bibinfo{volume}{64}},
  \bibinfo{pages}{51} (\bibinfo{year}{2014}{\natexlab{a}}).

\bibitem[{\citenamefont{Zhu et~al.}(2014{\natexlab{b}})\citenamefont{Zhu, Wu,
  Gong, Xiao, Li, Jin, Yao, Qian, Wu, Feng et~al.}}]{zhu2014emergence}
\bibinfo{author}{\bibfnamefont{K.}~\bibnamefont{Zhu}},
  \bibinfo{author}{\bibfnamefont{L.}~\bibnamefont{Wu}},
  \bibinfo{author}{\bibfnamefont{X.}~\bibnamefont{Gong}},
  \bibinfo{author}{\bibfnamefont{S.}~\bibnamefont{Xiao}},
  \bibinfo{author}{\bibfnamefont{S.}~\bibnamefont{Li}},
  \bibinfo{author}{\bibfnamefont{X.}~\bibnamefont{Jin}},
  \bibinfo{author}{\bibfnamefont{M.}~\bibnamefont{Yao}},
  \bibinfo{author}{\bibfnamefont{D.}~\bibnamefont{Qian}},
  \bibinfo{author}{\bibfnamefont{M.}~\bibnamefont{Wu}},
  \bibinfo{author}{\bibfnamefont{J.}~\bibnamefont{Feng}}, \bibnamefont{et~al.},
  \bibinfo{journal}{arXiv preprint arXiv:1403.0066}
  (\bibinfo{year}{2014}{\natexlab{b}}).

\bibitem[{\citenamefont{Hong et~al.}(2014)\citenamefont{Hong, Zhang, Cha, Qi,
  and Cui}}]{hong2014one}
\bibinfo{author}{\bibfnamefont{S.~S.} \bibnamefont{Hong}},
  \bibinfo{author}{\bibfnamefont{Y.}~\bibnamefont{Zhang}},
  \bibinfo{author}{\bibfnamefont{J.~J.} \bibnamefont{Cha}},
  \bibinfo{author}{\bibfnamefont{X.-L.} \bibnamefont{Qi}}, \bibnamefont{and}
  \bibinfo{author}{\bibfnamefont{Y.}~\bibnamefont{Cui}}, \bibinfo{journal}{Nano
  Lett.} \textbf{\bibinfo{volume}{14}}, \bibinfo{pages}{2815}
  (\bibinfo{year}{2014}).

\bibitem[{\citenamefont{Dufouleur et~al.}(2015)\citenamefont{Dufouleur, Veyrat,
  Xypakis, Bardarson, Nowka, Hampel, Eichler, Schmidt, B{\"u}chner, and
  Giraud}}]{dufouleur2015pseudo}
\bibinfo{author}{\bibfnamefont{J.}~\bibnamefont{Dufouleur}},
  \bibinfo{author}{\bibfnamefont{L.}~\bibnamefont{Veyrat}},
  \bibinfo{author}{\bibfnamefont{E.}~\bibnamefont{Xypakis}},
  \bibinfo{author}{\bibfnamefont{J.~H.} \bibnamefont{Bardarson}},
  \bibinfo{author}{\bibfnamefont{C.}~\bibnamefont{Nowka}},
  \bibinfo{author}{\bibfnamefont{S.}~\bibnamefont{Hampel}},
  \bibinfo{author}{\bibfnamefont{B.}~\bibnamefont{Eichler}},
  \bibinfo{author}{\bibfnamefont{O.~G.} \bibnamefont{Schmidt}},
  \bibinfo{author}{\bibfnamefont{B.}~\bibnamefont{B{\"u}chner}},
  \bibnamefont{and} \bibinfo{author}{\bibfnamefont{R.}~\bibnamefont{Giraud}},
  \bibinfo{journal}{arXiv preprint arXiv:1504.08030}  (\bibinfo{year}{2015}).

\bibitem[{\citenamefont{Jauregui et~al.}(2016)\citenamefont{Jauregui, Pettes,
  Rokhinson, Shi, and Chen}}]{jauregui2016magnetic}
\bibinfo{author}{\bibfnamefont{L.~A.} \bibnamefont{Jauregui}},
  \bibinfo{author}{\bibfnamefont{M.~T.} \bibnamefont{Pettes}},
  \bibinfo{author}{\bibfnamefont{L.~P.} \bibnamefont{Rokhinson}},
  \bibinfo{author}{\bibfnamefont{L.}~\bibnamefont{Shi}}, \bibnamefont{and}
  \bibinfo{author}{\bibfnamefont{Y.~P.} \bibnamefont{Chen}},
  \bibinfo{journal}{Nat. Nano.} p. \bibinfo{pages}{345} (\bibinfo{year}{2016}).

\bibitem[{\citenamefont{Jauregui et~al.}(2015)\citenamefont{Jauregui, Pettes,
  Rokhinson, Shi, and Chen}}]{jauregui2015gate}
\bibinfo{author}{\bibfnamefont{L.~A.} \bibnamefont{Jauregui}},
  \bibinfo{author}{\bibfnamefont{M.~T.} \bibnamefont{Pettes}},
  \bibinfo{author}{\bibfnamefont{L.~P.} \bibnamefont{Rokhinson}},
  \bibinfo{author}{\bibfnamefont{L.}~\bibnamefont{Shi}}, \bibnamefont{and}
  \bibinfo{author}{\bibfnamefont{Y.~P.} \bibnamefont{Chen}},
  \bibinfo{journal}{Sci. Rep.} \textbf{\bibinfo{volume}{5}}
  (\bibinfo{year}{2015}).

\bibitem[{\citenamefont{Cho et~al.}(2015)\citenamefont{Cho, Dellabetta, Zhong,
  Schneeloch, Liu, Gu, Gilbert, and Mason}}]{cho2015aharonov}
\bibinfo{author}{\bibfnamefont{S.}~\bibnamefont{Cho}},
  \bibinfo{author}{\bibfnamefont{B.}~\bibnamefont{Dellabetta}},
  \bibinfo{author}{\bibfnamefont{R.}~\bibnamefont{Zhong}},
  \bibinfo{author}{\bibfnamefont{J.}~\bibnamefont{Schneeloch}},
  \bibinfo{author}{\bibfnamefont{T.}~\bibnamefont{Liu}},
  \bibinfo{author}{\bibfnamefont{G.}~\bibnamefont{Gu}},
  \bibinfo{author}{\bibfnamefont{M.~J.} \bibnamefont{Gilbert}},
  \bibnamefont{and} \bibinfo{author}{\bibfnamefont{N.}~\bibnamefont{Mason}},
  \bibinfo{journal}{Nat. Comm.} \textbf{\bibinfo{volume}{6}}
  (\bibinfo{year}{2015}).

\bibitem[{\citenamefont{Kim et~al.}(2016{\natexlab{a}})\citenamefont{Kim, Shin,
  Lee, Ahn, Song, and Doh}}]{kim2016quantum}
\bibinfo{author}{\bibfnamefont{H.-S.} \bibnamefont{Kim}},
  \bibinfo{author}{\bibfnamefont{H.~S.} \bibnamefont{Shin}},
  \bibinfo{author}{\bibfnamefont{J.~S.} \bibnamefont{Lee}},
  \bibinfo{author}{\bibfnamefont{C.~W.} \bibnamefont{Ahn}},
  \bibinfo{author}{\bibfnamefont{J.~Y.} \bibnamefont{Song}}, \bibnamefont{and}
  \bibinfo{author}{\bibfnamefont{Y.-J.} \bibnamefont{Doh}},
  \bibinfo{journal}{Curr. App. Phys.} \textbf{\bibinfo{volume}{16}},
  \bibinfo{pages}{51} (\bibinfo{year}{2016}{\natexlab{a}}).

\bibitem[{\citenamefont{Kim et~al.}(2016{\natexlab{b}})\citenamefont{Kim,
  Hwang, Lee, Jhi, Lee, Park, Kim, Kim, Doh, Kim et~al.}}]{kim2016quantumbeta}
\bibinfo{author}{\bibfnamefont{J.}~\bibnamefont{Kim}},
  \bibinfo{author}{\bibfnamefont{A.}~\bibnamefont{Hwang}},
  \bibinfo{author}{\bibfnamefont{S.-H.} \bibnamefont{Lee}},
  \bibinfo{author}{\bibfnamefont{S.-H.} \bibnamefont{Jhi}},
  \bibinfo{author}{\bibfnamefont{S.}~\bibnamefont{Lee}},
  \bibinfo{author}{\bibfnamefont{Y.~C.} \bibnamefont{Park}},
  \bibinfo{author}{\bibfnamefont{S.-i.} \bibnamefont{Kim}},
  \bibinfo{author}{\bibfnamefont{H.-S.} \bibnamefont{Kim}},
  \bibinfo{author}{\bibfnamefont{Y.-J.} \bibnamefont{Doh}},
  \bibinfo{author}{\bibfnamefont{J.}~\bibnamefont{Kim}}, \bibnamefont{et~al.},
  \bibinfo{journal}{arXiv preprint arXiv:1601.01551}
  (\bibinfo{year}{2016}{\natexlab{b}}).

\bibitem[{\citenamefont{Checkelsky et~al.}(2011)\citenamefont{Checkelsky, Hor,
  Cava, and Ong}}]{PhysRevLett.106.196801}
\bibinfo{author}{\bibfnamefont{J.~G.} \bibnamefont{Checkelsky}},
  \bibinfo{author}{\bibfnamefont{Y.~S.} \bibnamefont{Hor}},
  \bibinfo{author}{\bibfnamefont{R.~J.} \bibnamefont{Cava}}, \bibnamefont{and}
  \bibinfo{author}{\bibfnamefont{N.~P.} \bibnamefont{Ong}},
  \bibinfo{journal}{Phys. Rev. Lett.} \textbf{\bibinfo{volume}{106}},
  \bibinfo{pages}{196801} (\bibinfo{year}{2011}).

\bibitem[{\citenamefont{Matsuo et~al.}(2012)\citenamefont{Matsuo, Koyama,
  Shimamura, Arakawa, Nishihara, Chiba, Kobayashi, Ono, Chang, He
  et~al.}}]{PhysRevB.85.075440}
\bibinfo{author}{\bibfnamefont{S.}~\bibnamefont{Matsuo}},
  \bibinfo{author}{\bibfnamefont{T.}~\bibnamefont{Koyama}},
  \bibinfo{author}{\bibfnamefont{K.}~\bibnamefont{Shimamura}},
  \bibinfo{author}{\bibfnamefont{T.}~\bibnamefont{Arakawa}},
  \bibinfo{author}{\bibfnamefont{Y.}~\bibnamefont{Nishihara}},
  \bibinfo{author}{\bibfnamefont{D.}~\bibnamefont{Chiba}},
  \bibinfo{author}{\bibfnamefont{K.}~\bibnamefont{Kobayashi}},
  \bibinfo{author}{\bibfnamefont{T.}~\bibnamefont{Ono}},
  \bibinfo{author}{\bibfnamefont{C.-Z.} \bibnamefont{Chang}},
  \bibinfo{author}{\bibfnamefont{K.}~\bibnamefont{He}}, \bibnamefont{et~al.},
  \bibinfo{journal}{Phys. Rev. B} \textbf{\bibinfo{volume}{85}},
  \bibinfo{pages}{075440} (\bibinfo{year}{2012}).

\bibitem[{\citenamefont{Matsuo et~al.}(2013)\citenamefont{Matsuo, Chida, Chiba,
  Ono, Slevin, Kobayashi, Ohtsuki, Chang, He, Ma et~al.}}]{PhysRevB.88.155438}
\bibinfo{author}{\bibfnamefont{S.}~\bibnamefont{Matsuo}},
  \bibinfo{author}{\bibfnamefont{K.}~\bibnamefont{Chida}},
  \bibinfo{author}{\bibfnamefont{D.}~\bibnamefont{Chiba}},
  \bibinfo{author}{\bibfnamefont{T.}~\bibnamefont{Ono}},
  \bibinfo{author}{\bibfnamefont{K.}~\bibnamefont{Slevin}},
  \bibinfo{author}{\bibfnamefont{K.}~\bibnamefont{Kobayashi}},
  \bibinfo{author}{\bibfnamefont{T.}~\bibnamefont{Ohtsuki}},
  \bibinfo{author}{\bibfnamefont{C.-Z.} \bibnamefont{Chang}},
  \bibinfo{author}{\bibfnamefont{K.}~\bibnamefont{He}},
  \bibinfo{author}{\bibfnamefont{X.-C.} \bibnamefont{Ma}},
  \bibnamefont{et~al.}, \bibinfo{journal}{Phys. Rev. B}
  \textbf{\bibinfo{volume}{88}}, \bibinfo{pages}{155438}
  (\bibinfo{year}{2013}).

\bibitem[{\citenamefont{Zhao et~al.}(2015)\citenamefont{Zhao, Chen, Pan, Fei,
  and Han}}]{zhao2015electronic}
\bibinfo{author}{\bibfnamefont{B.}~\bibnamefont{Zhao}},
  \bibinfo{author}{\bibfnamefont{T.}~\bibnamefont{Chen}},
  \bibinfo{author}{\bibfnamefont{H.}~\bibnamefont{Pan}},
  \bibinfo{author}{\bibfnamefont{F.}~\bibnamefont{Fei}}, \bibnamefont{and}
  \bibinfo{author}{\bibfnamefont{Y.}~\bibnamefont{Han}}, \bibinfo{journal}{J.
  Phys.: Cond. Matt.} \textbf{\bibinfo{volume}{27}}, \bibinfo{pages}{465302}
  (\bibinfo{year}{2015}).

\bibitem[{\citenamefont{Xia et~al.}(2013)\citenamefont{Xia, Ren, Sulaev, Liu,
  Shen, and Wang}}]{PhysRevB.87.085442}
\bibinfo{author}{\bibfnamefont{B.}~\bibnamefont{Xia}},
  \bibinfo{author}{\bibfnamefont{P.}~\bibnamefont{Ren}},
  \bibinfo{author}{\bibfnamefont{A.}~\bibnamefont{Sulaev}},
  \bibinfo{author}{\bibfnamefont{P.}~\bibnamefont{Liu}},
  \bibinfo{author}{\bibfnamefont{S.-Q.} \bibnamefont{Shen}}, \bibnamefont{and}
  \bibinfo{author}{\bibfnamefont{L.}~\bibnamefont{Wang}},
  \bibinfo{journal}{Phys. Rev. B} \textbf{\bibinfo{volume}{87}},
  \bibinfo{pages}{085442} (\bibinfo{year}{2013}).

\bibitem[{\citenamefont{Li et~al.}(2014)\citenamefont{Li, Meng, Pan, Chen,
  Hong, Li, Wang, Song, and Wang}}]{li2014indications}
\bibinfo{author}{\bibfnamefont{Z.}~\bibnamefont{Li}},
  \bibinfo{author}{\bibfnamefont{Y.}~\bibnamefont{Meng}},
  \bibinfo{author}{\bibfnamefont{J.}~\bibnamefont{Pan}},
  \bibinfo{author}{\bibfnamefont{T.}~\bibnamefont{Chen}},
  \bibinfo{author}{\bibfnamefont{X.}~\bibnamefont{Hong}},
  \bibinfo{author}{\bibfnamefont{S.}~\bibnamefont{Li}},
  \bibinfo{author}{\bibfnamefont{X.}~\bibnamefont{Wang}},
  \bibinfo{author}{\bibfnamefont{F.}~\bibnamefont{Song}}, \bibnamefont{and}
  \bibinfo{author}{\bibfnamefont{B.}~\bibnamefont{Wang}},
  \bibinfo{journal}{Appl. Phys. Expr.} \textbf{\bibinfo{volume}{7}},
  \bibinfo{pages}{065202} (\bibinfo{year}{2014}).

\bibitem[{\citenamefont{Kim et~al.}(2014)\citenamefont{Kim, Lee, Brovman, Kim,
  Kim, and Lee}}]{kim2014weak}
\bibinfo{author}{\bibfnamefont{J.}~\bibnamefont{Kim}},
  \bibinfo{author}{\bibfnamefont{S.}~\bibnamefont{Lee}},
  \bibinfo{author}{\bibfnamefont{Y.~M.} \bibnamefont{Brovman}},
  \bibinfo{author}{\bibfnamefont{M.}~\bibnamefont{Kim}},
  \bibinfo{author}{\bibfnamefont{P.}~\bibnamefont{Kim}}, \bibnamefont{and}
  \bibinfo{author}{\bibfnamefont{W.}~\bibnamefont{Lee}},
  \bibinfo{journal}{Appl. Phys. Lett.} \textbf{\bibinfo{volume}{104}},
  \bibinfo{pages}{043105} (\bibinfo{year}{2014}).

\bibitem[{\citenamefont{Zirnbauer}(1992)}]{PhysRevLett.69.1584}
\bibinfo{author}{\bibfnamefont{M.~R.} \bibnamefont{Zirnbauer}},
  \bibinfo{journal}{Phys. Rev. Lett.} \textbf{\bibinfo{volume}{69}},
  \bibinfo{pages}{1584} (\bibinfo{year}{1992}).

\bibitem[{\citenamefont{Mirlin et~al.}(1994)\citenamefont{Mirlin,
  Muller-Groeling, and Zirnbauer}}]{Mirlin94}
\bibinfo{author}{\bibfnamefont{A.~D.} \bibnamefont{Mirlin}},
  \bibinfo{author}{\bibfnamefont{A.}~\bibnamefont{Muller-Groeling}},
  \bibnamefont{and} \bibinfo{author}{\bibfnamefont{M.~R.}
  \bibnamefont{Zirnbauer}}, \bibinfo{journal}{Ann. Phys.}
  \textbf{\bibinfo{volume}{236}}, \bibinfo{pages}{325} (\bibinfo{year}{1994}).

\bibitem[{\citenamefont{Ando and Suzuura}(2002)}]{Ando02}
\bibinfo{author}{\bibfnamefont{T.}~\bibnamefont{Ando}} \bibnamefont{and}
  \bibinfo{author}{\bibfnamefont{H.}~\bibnamefont{Suzuura}},
  \bibinfo{journal}{J. Phys. Soc. Jpn.} \textbf{\bibinfo{volume}{71}},
  \bibinfo{pages}{2753} (\bibinfo{year}{2002}).

\bibitem[{\citenamefont{Takane}(2004)}]{Takane04}
\bibinfo{author}{\bibfnamefont{Y.}~\bibnamefont{Takane}}, \bibinfo{journal}{J.
  Phys. Soc. Jpn.} \textbf{\bibinfo{volume}{73}}, \bibinfo{pages}{1430}
  (\bibinfo{year}{2004}).

\bibitem[{\citenamefont{Ryu et~al.}(2007)\citenamefont{Ryu, Mudry, Obuse, and
  Furusaki}}]{PhysRevLett.99.116601}
\bibinfo{author}{\bibfnamefont{S.}~\bibnamefont{Ryu}},
  \bibinfo{author}{\bibfnamefont{C.}~\bibnamefont{Mudry}},
  \bibinfo{author}{\bibfnamefont{H.}~\bibnamefont{Obuse}}, \bibnamefont{and}
  \bibinfo{author}{\bibfnamefont{A.}~\bibnamefont{Furusaki}},
  \bibinfo{journal}{Phys. Rev. Lett.} \textbf{\bibinfo{volume}{99}},
  \bibinfo{pages}{116601} (\bibinfo{year}{2007}).

\bibitem[{\citenamefont{Wakabayashi et~al.}(2007)\citenamefont{Wakabayashi,
  Takane, and Sigrist}}]{PhysRevLett.99.036601}
\bibinfo{author}{\bibfnamefont{K.}~\bibnamefont{Wakabayashi}},
  \bibinfo{author}{\bibfnamefont{Y.}~\bibnamefont{Takane}}, \bibnamefont{and}
  \bibinfo{author}{\bibfnamefont{M.}~\bibnamefont{Sigrist}},
  \bibinfo{journal}{Phys. Rev. Lett.} \textbf{\bibinfo{volume}{99}},
  \bibinfo{pages}{036601} (\bibinfo{year}{2007}).

\bibitem[{\citenamefont{Ashitani et~al.}(2012)\citenamefont{Ashitani, Imura,
  and Takane}}]{ashitani2012perfectly}
\bibinfo{author}{\bibfnamefont{Y.}~\bibnamefont{Ashitani}},
  \bibinfo{author}{\bibfnamefont{K.-I.} \bibnamefont{Imura}}, \bibnamefont{and}
  \bibinfo{author}{\bibfnamefont{Y.}~\bibnamefont{Takane}}, in
  \emph{\bibinfo{booktitle}{International Journal of Modern Physics: Conference
  Series}} (\bibinfo{organization}{World Scientific}, \bibinfo{year}{2012}),
  vol.~\bibinfo{volume}{11}, pp. \bibinfo{pages}{157--162}.

\bibitem[{\citenamefont{Wu and Sacksteder}(2014)}]{Wu14}
\bibinfo{author}{\bibfnamefont{Q.}~\bibnamefont{Wu}} \bibnamefont{and}
  \bibinfo{author}{\bibfnamefont{V.~E.} \bibnamefont{Sacksteder}},
  \bibinfo{journal}{Phys. Rev. B} \textbf{\bibinfo{volume}{90}},
  \bibinfo{pages}{045408} (\bibinfo{year}{2014}).

\bibitem[{\citenamefont{Kolomeisky and Straley}(2014)}]{kolomeisky2014aharonov}
\bibinfo{author}{\bibfnamefont{E.~B.} \bibnamefont{Kolomeisky}}
  \bibnamefont{and} \bibinfo{author}{\bibfnamefont{J.~P.}
  \bibnamefont{Straley}}, \bibinfo{journal}{arXiv preprint arXiv:1409.7974}
  (\bibinfo{year}{2014}).

\bibitem[{\citenamefont{Shimomura and Takane}(2015)}]{shimomura2015dephasing}
\bibinfo{author}{\bibfnamefont{Y.}~\bibnamefont{Shimomura}} \bibnamefont{and}
  \bibinfo{author}{\bibfnamefont{Y.}~\bibnamefont{Takane}},
  \bibinfo{journal}{J. Phys. Soc. Jpn.} \textbf{\bibinfo{volume}{85}},
  \bibinfo{pages}{014704} (\bibinfo{year}{2015}).

\bibitem[{\citenamefont{Zhang and Vishwanath}(2010)}]{PhysRevLett.105.206601}
\bibinfo{author}{\bibfnamefont{Y.}~\bibnamefont{Zhang}} \bibnamefont{and}
  \bibinfo{author}{\bibfnamefont{A.}~\bibnamefont{Vishwanath}},
  \bibinfo{journal}{Phys. Rev. Lett.} \textbf{\bibinfo{volume}{105}},
  \bibinfo{pages}{206601} (\bibinfo{year}{2010}).

\bibitem[{\citenamefont{Egger et~al.}(2010)\citenamefont{Egger, Zazunov, and
  Yeyati}}]{PhysRevLett.105.136403}
\bibinfo{author}{\bibfnamefont{R.}~\bibnamefont{Egger}},
  \bibinfo{author}{\bibfnamefont{A.}~\bibnamefont{Zazunov}}, \bibnamefont{and}
  \bibinfo{author}{\bibfnamefont{A.~L.} \bibnamefont{Yeyati}},
  \bibinfo{journal}{Phys. Rev. Lett.} \textbf{\bibinfo{volume}{105}},
  \bibinfo{pages}{136403} (\bibinfo{year}{2010}).

\bibitem[{\citenamefont{Lee and Ramakrishnan}(1985)}]{RevModPhys.57.287}
\bibinfo{author}{\bibfnamefont{P.~A.} \bibnamefont{Lee}} \bibnamefont{and}
  \bibinfo{author}{\bibfnamefont{T.~V.} \bibnamefont{Ramakrishnan}},
  \bibinfo{journal}{Rev. Mod. Phys.} \textbf{\bibinfo{volume}{57}},
  \bibinfo{pages}{287} (\bibinfo{year}{1985}).

\bibitem[{\citenamefont{Rammer and Smith}(1986)}]{RevModPhys.58.323}
\bibinfo{author}{\bibfnamefont{J.}~\bibnamefont{Rammer}} \bibnamefont{and}
  \bibinfo{author}{\bibfnamefont{H.}~\bibnamefont{Smith}},
  \bibinfo{journal}{Rev. Mod. Phys.} \textbf{\bibinfo{volume}{58}},
  \bibinfo{pages}{323} (\bibinfo{year}{1986}).

\bibitem[{\citenamefont{Beenakker}(1997)}]{RevModPhys.69.731}
\bibinfo{author}{\bibfnamefont{C.~W.~J.} \bibnamefont{Beenakker}},
  \bibinfo{journal}{Rev. Mod. Phys.} \textbf{\bibinfo{volume}{69}},
  \bibinfo{pages}{731} (\bibinfo{year}{1997}).

\bibitem[{\citenamefont{Bardarson et~al.}(2010)\citenamefont{Bardarson,
  Brouwer, and Moore}}]{PhysRevLett.105.156803}
\bibinfo{author}{\bibfnamefont{J.~H.} \bibnamefont{Bardarson}},
  \bibinfo{author}{\bibfnamefont{P.~W.} \bibnamefont{Brouwer}},
  \bibnamefont{and} \bibinfo{author}{\bibfnamefont{J.~E.} \bibnamefont{Moore}},
  \bibinfo{journal}{Phys. Rev. Lett.} \textbf{\bibinfo{volume}{105}},
  \bibinfo{pages}{156803} (\bibinfo{year}{2010}).

\bibitem[{\citenamefont{Bardarson and Moore}(2013)}]{bardarson2013quantum}
\bibinfo{author}{\bibfnamefont{J.~H.} \bibnamefont{Bardarson}}
  \bibnamefont{and} \bibinfo{author}{\bibfnamefont{J.~E.} \bibnamefont{Moore}},
  \bibinfo{journal}{Rep. Progr. Phys.} \textbf{\bibinfo{volume}{76}},
  \bibinfo{pages}{056501} (\bibinfo{year}{2013}).

\bibitem[{\citenamefont{Adroguer et~al.}(2012)\citenamefont{Adroguer,
  Carpentier, Cayssol, and Orignac}}]{adroguer2012diffusion}
\bibinfo{author}{\bibfnamefont{P.}~\bibnamefont{Adroguer}},
  \bibinfo{author}{\bibfnamefont{D.}~\bibnamefont{Carpentier}},
  \bibinfo{author}{\bibfnamefont{J.}~\bibnamefont{Cayssol}}, \bibnamefont{and}
  \bibinfo{author}{\bibfnamefont{E.}~\bibnamefont{Orignac}},
  \bibinfo{journal}{New J. Phys.} \textbf{\bibinfo{volume}{14}},
  \bibinfo{pages}{103027} (\bibinfo{year}{2012}).

\bibitem[{\citenamefont{Chen et~al.}(2010)\citenamefont{Chen, Qin, Yang, Liu,
  Guan, Qu, Zhang, Shi, Xie, Yang et~al.}}]{PhysRevLett.105.176602}
\bibinfo{author}{\bibfnamefont{J.}~\bibnamefont{Chen}},
  \bibinfo{author}{\bibfnamefont{H.~J.} \bibnamefont{Qin}},
  \bibinfo{author}{\bibfnamefont{F.}~\bibnamefont{Yang}},
  \bibinfo{author}{\bibfnamefont{J.}~\bibnamefont{Liu}},
  \bibinfo{author}{\bibfnamefont{T.}~\bibnamefont{Guan}},
  \bibinfo{author}{\bibfnamefont{F.~M.} \bibnamefont{Qu}},
  \bibinfo{author}{\bibfnamefont{G.~H.} \bibnamefont{Zhang}},
  \bibinfo{author}{\bibfnamefont{J.~R.} \bibnamefont{Shi}},
  \bibinfo{author}{\bibfnamefont{X.~C.} \bibnamefont{Xie}},
  \bibinfo{author}{\bibfnamefont{C.~L.} \bibnamefont{Yang}},
  \bibnamefont{et~al.}, \bibinfo{journal}{Phys. Rev. Lett.}
  \textbf{\bibinfo{volume}{105}}, \bibinfo{pages}{176602}
  (\bibinfo{year}{2010}).

\bibitem[{\citenamefont{Wen-Min et~al.}(2013)\citenamefont{Wen-Min, Chao-Jing,
  Jian, and Yong-Qing}}]{wen2013electrostatic}
\bibinfo{author}{\bibfnamefont{Y.}~\bibnamefont{Wen-Min}},
  \bibinfo{author}{\bibfnamefont{L.}~\bibnamefont{Chao-Jing}},
  \bibinfo{author}{\bibfnamefont{L.}~\bibnamefont{Jian}}, \bibnamefont{and}
  \bibinfo{author}{\bibfnamefont{L.}~\bibnamefont{Yong-Qing}},
  \bibinfo{journal}{Chin. Phys. B} \textbf{\bibinfo{volume}{22}},
  \bibinfo{pages}{097202} (\bibinfo{year}{2013}).

\bibitem[{\citenamefont{Zhang et~al.}(2010)\citenamefont{Zhang, Yu, Zhang, Dai,
  and Fang}}]{zhang2010first}
\bibinfo{author}{\bibfnamefont{W.}~\bibnamefont{Zhang}},
  \bibinfo{author}{\bibfnamefont{R.}~\bibnamefont{Yu}},
  \bibinfo{author}{\bibfnamefont{H.-J.} \bibnamefont{Zhang}},
  \bibinfo{author}{\bibfnamefont{X.}~\bibnamefont{Dai}}, \bibnamefont{and}
  \bibinfo{author}{\bibfnamefont{Z.}~\bibnamefont{Fang}}, \bibinfo{journal}{New
  Journal of Physics} \textbf{\bibinfo{volume}{12}}, \bibinfo{pages}{065013}
  (\bibinfo{year}{2010}).

\bibitem[{\citenamefont{Liu et~al.}(2010)\citenamefont{Liu, Qi, Zhang, Dai,
  Fang, and Zhang}}]{liu2010model}
\bibinfo{author}{\bibfnamefont{C.-X.} \bibnamefont{Liu}},
  \bibinfo{author}{\bibfnamefont{X.-L.} \bibnamefont{Qi}},
  \bibinfo{author}{\bibfnamefont{H.~J.} \bibnamefont{Zhang}},
  \bibinfo{author}{\bibfnamefont{X.}~\bibnamefont{Dai}},
  \bibinfo{author}{\bibfnamefont{Z.}~\bibnamefont{Fang}}, \bibnamefont{and}
  \bibinfo{author}{\bibfnamefont{S.-C.} \bibnamefont{Zhang}},
  \bibinfo{journal}{Phys. Rev. B} \textbf{\bibinfo{volume}{82}},
  \bibinfo{pages}{045122} (\bibinfo{year}{2010}).

\bibitem[{\citenamefont{Ryu and Nomura}(2012)}]{ryu2012disorder}
\bibinfo{author}{\bibfnamefont{S.}~\bibnamefont{Ryu}} \bibnamefont{and}
  \bibinfo{author}{\bibfnamefont{K.}~\bibnamefont{Nomura}},
  \bibinfo{journal}{Phys. Rev. B} \textbf{\bibinfo{volume}{85}},
  \bibinfo{pages}{155138} (\bibinfo{year}{2012}).

\bibitem[{\citenamefont{Kobayashi et~al.}(2013)\citenamefont{Kobayashi,
  Ohtsuki, and Imura}}]{kobayashi2013disordered}
\bibinfo{author}{\bibfnamefont{K.}~\bibnamefont{Kobayashi}},
  \bibinfo{author}{\bibfnamefont{T.}~\bibnamefont{Ohtsuki}}, \bibnamefont{and}
  \bibinfo{author}{\bibfnamefont{K.-I.} \bibnamefont{Imura}},
  \bibinfo{journal}{Phys. Rev. Lett.} \textbf{\bibinfo{volume}{110}},
  \bibinfo{pages}{236803} (\bibinfo{year}{2013}).

\bibitem[{\citenamefont{Kobayashi et~al.}(2014)\citenamefont{Kobayashi,
  Ohtsuki, Imura, and Herbut}}]{kobayashi2014density}
\bibinfo{author}{\bibfnamefont{K.}~\bibnamefont{Kobayashi}},
  \bibinfo{author}{\bibfnamefont{T.}~\bibnamefont{Ohtsuki}},
  \bibinfo{author}{\bibfnamefont{K.-I.} \bibnamefont{Imura}}, \bibnamefont{and}
  \bibinfo{author}{\bibfnamefont{I.~F.} \bibnamefont{Herbut}},
  \bibinfo{journal}{Phys. Rev. Lett.} \textbf{\bibinfo{volume}{112}},
  \bibinfo{pages}{016402} (\bibinfo{year}{2014}).

\bibitem[{\citenamefont{Sacksteder et~al.}(2015)\citenamefont{Sacksteder,
  Ohtsuki, and Kobayashi}}]{PhysRevApplied.3.064006}
\bibinfo{author}{\bibfnamefont{V.}~\bibnamefont{Sacksteder}},
  \bibinfo{author}{\bibfnamefont{T.}~\bibnamefont{Ohtsuki}}, \bibnamefont{and}
  \bibinfo{author}{\bibfnamefont{K.}~\bibnamefont{Kobayashi}},
  \bibinfo{journal}{Phys. Rev. Applied} \textbf{\bibinfo{volume}{3}},
  \bibinfo{pages}{064006} (\bibinfo{year}{2015}).

\bibitem[{\citenamefont{Caroli et~al.}(1971)\citenamefont{Caroli, Combescot,
  Nozieres, and Saint-James}}]{Caroli71}
\bibinfo{author}{\bibfnamefont{C.}~\bibnamefont{Caroli}},
  \bibinfo{author}{\bibfnamefont{R.}~\bibnamefont{Combescot}},
  \bibinfo{author}{\bibfnamefont{P.}~\bibnamefont{Nozieres}}, \bibnamefont{and}
  \bibinfo{author}{\bibfnamefont{D.}~\bibnamefont{Saint-James}},
  \bibinfo{journal}{Journal of Physics C} \textbf{\bibinfo{volume}{4}},
  \bibinfo{pages}{916} (\bibinfo{year}{1971}).

\bibitem[{\citenamefont{Meir and Wingreen}(1992)}]{Meir92}
\bibinfo{author}{\bibfnamefont{Y.}~\bibnamefont{Meir}} \bibnamefont{and}
  \bibinfo{author}{\bibfnamefont{N.~S.} \bibnamefont{Wingreen}},
  \bibinfo{journal}{Phys. Rev. Lett.} \textbf{\bibinfo{volume}{68}},
  \bibinfo{pages}{2512} (\bibinfo{year}{1992}).

\bibitem[{\citenamefont{Sancho et~al.}(1985)\citenamefont{Sancho, Sancho, and
  Rubio}}]{Lopez84}
\bibinfo{author}{\bibfnamefont{M.~P.~L.} \bibnamefont{Sancho}},
  \bibinfo{author}{\bibfnamefont{J.~M.~L.} \bibnamefont{Sancho}},
  \bibnamefont{and} \bibinfo{author}{\bibfnamefont{J.}~\bibnamefont{Rubio}},
  \bibinfo{journal}{Journal of Physics F: Metal Physics}
  \textbf{\bibinfo{volume}{15}}, \bibinfo{pages}{851} (\bibinfo{year}{1985}).

\bibitem[{\citenamefont{Lewenkopf and Mucciolo}(2013)}]{lewenkopf2013recursive}
\bibinfo{author}{\bibfnamefont{C.~H.} \bibnamefont{Lewenkopf}}
  \bibnamefont{and} \bibinfo{author}{\bibfnamefont{E.~R.}
  \bibnamefont{Mucciolo}}, \bibinfo{journal}{Journal of Computational
  Electronics} \textbf{\bibinfo{volume}{12}}, \bibinfo{pages}{203}
  (\bibinfo{year}{2013}).

\bibitem[{\citenamefont{Lin et~al.}(2013)\citenamefont{Lin, He, Liao, Wang, IV,
  Yang, Guan, Zhang, Gu, Zhang et~al.}}]{PhysRevB.88.041307}
\bibinfo{author}{\bibfnamefont{C.~J.} \bibnamefont{Lin}},
  \bibinfo{author}{\bibfnamefont{X.~Y.} \bibnamefont{He}},
  \bibinfo{author}{\bibfnamefont{J.}~\bibnamefont{Liao}},
  \bibinfo{author}{\bibfnamefont{X.~X.} \bibnamefont{Wang}},
  \bibinfo{author}{\bibfnamefont{V.~S.} \bibnamefont{IV}},
  \bibinfo{author}{\bibfnamefont{W.~M.} \bibnamefont{Yang}},
  \bibinfo{author}{\bibfnamefont{T.}~\bibnamefont{Guan}},
  \bibinfo{author}{\bibfnamefont{Q.~M.} \bibnamefont{Zhang}},
  \bibinfo{author}{\bibfnamefont{L.}~\bibnamefont{Gu}},
  \bibinfo{author}{\bibfnamefont{G.~Y.} \bibnamefont{Zhang}},
  \bibnamefont{et~al.}, \bibinfo{journal}{Phys. Rev. B}
  \textbf{\bibinfo{volume}{88}}, \bibinfo{pages}{041307}
  (\bibinfo{year}{2013}).

\bibitem[{\citenamefont{Kawabata and Nakamura}(1996)}]{kawabata1996altshuler}
\bibinfo{author}{\bibfnamefont{S.}~\bibnamefont{Kawabata}} \bibnamefont{and}
  \bibinfo{author}{\bibfnamefont{K.}~\bibnamefont{Nakamura}},
  \bibinfo{journal}{J. Phys. Soc. Jpn.} \textbf{\bibinfo{volume}{65}},
  \bibinfo{pages}{3708} (\bibinfo{year}{1996}).

\bibitem[{\citenamefont{Kawabata and Nakamura}(1998)}]{PhysRevB.57.6282}
\bibinfo{author}{\bibfnamefont{S.}~\bibnamefont{Kawabata}} \bibnamefont{and}
  \bibinfo{author}{\bibfnamefont{K.}~\bibnamefont{Nakamura}},
  \bibinfo{journal}{Phys. Rev. B} \textbf{\bibinfo{volume}{57}},
  \bibinfo{pages}{6282} (\bibinfo{year}{1998}).

\bibitem[{\citenamefont{Lee et~al.}(2004)\citenamefont{Lee, Cho, Ihm, and
  Ahn}}]{PhysRevB.69.205403}
\bibinfo{author}{\bibfnamefont{C.-K.} \bibnamefont{Lee}},
  \bibinfo{author}{\bibfnamefont{J.}~\bibnamefont{Cho}},
  \bibinfo{author}{\bibfnamefont{J.}~\bibnamefont{Ihm}}, \bibnamefont{and}
  \bibinfo{author}{\bibfnamefont{K.-H.} \bibnamefont{Ahn}},
  \bibinfo{journal}{Phys. Rev. B} \textbf{\bibinfo{volume}{69}},
  \bibinfo{pages}{205403} (\bibinfo{year}{2004}).

\bibitem[{\citenamefont{Koga et~al.}(2004)\citenamefont{Koga, Nitta, and van
  Veenhuizen}}]{PhysRevB.70.161302}
\bibinfo{author}{\bibfnamefont{T.}~\bibnamefont{Koga}},
  \bibinfo{author}{\bibfnamefont{J.}~\bibnamefont{Nitta}}, \bibnamefont{and}
  \bibinfo{author}{\bibfnamefont{M.}~\bibnamefont{van Veenhuizen}},
  \bibinfo{journal}{Phys. Rev. B} \textbf{\bibinfo{volume}{70}},
  \bibinfo{pages}{161302} (\bibinfo{year}{2004}).

\bibitem[{\citenamefont{Washburn et~al.}(1985)\citenamefont{Washburn, Umbach,
  Laibowitz, and Webb}}]{PhysRevB.32.4789}
\bibinfo{author}{\bibfnamefont{S.}~\bibnamefont{Washburn}},
  \bibinfo{author}{\bibfnamefont{C.~P.} \bibnamefont{Umbach}},
  \bibinfo{author}{\bibfnamefont{R.~B.} \bibnamefont{Laibowitz}},
  \bibnamefont{and} \bibinfo{author}{\bibfnamefont{R.~A.} \bibnamefont{Webb}},
  \bibinfo{journal}{Phys. Rev. B} \textbf{\bibinfo{volume}{32}},
  \bibinfo{pages}{4789} (\bibinfo{year}{1985}).

\bibitem[{\citenamefont{Hansen et~al.}(2001)\citenamefont{Hansen, Kristensen,
  Pedersen, S\o{}rensen, and Lindelof}}]{PhysRevB.64.045327}
\bibinfo{author}{\bibfnamefont{A.~E.} \bibnamefont{Hansen}},
  \bibinfo{author}{\bibfnamefont{A.}~\bibnamefont{Kristensen}},
  \bibinfo{author}{\bibfnamefont{S.}~\bibnamefont{Pedersen}},
  \bibinfo{author}{\bibfnamefont{C.~B.} \bibnamefont{S\o{}rensen}},
  \bibnamefont{and} \bibinfo{author}{\bibfnamefont{P.~E.}
  \bibnamefont{Lindelof}}, \bibinfo{journal}{Phys. Rev. B}
  \textbf{\bibinfo{volume}{64}}, \bibinfo{pages}{045327}
  (\bibinfo{year}{2001}).

\bibitem[{\citenamefont{Takane}(2010)}]{takane2010anomalous}
\bibinfo{author}{\bibfnamefont{Y.}~\bibnamefont{Takane}}, \bibinfo{journal}{J.
  Phys. Soc. Jpn.} \textbf{\bibinfo{volume}{79}}, \bibinfo{pages}{024711}
  (\bibinfo{year}{2010}).

\bibitem[{\citenamefont{Schmidt et~al.}(2012)\citenamefont{Schmidt, Rachel, von
  Oppen, and Glazman}}]{PhysRevLett.108.156402}
\bibinfo{author}{\bibfnamefont{T.~L.} \bibnamefont{Schmidt}},
  \bibinfo{author}{\bibfnamefont{S.}~\bibnamefont{Rachel}},
  \bibinfo{author}{\bibfnamefont{F.}~\bibnamefont{von Oppen}},
  \bibnamefont{and} \bibinfo{author}{\bibfnamefont{L.~I.}
  \bibnamefont{Glazman}}, \bibinfo{journal}{Phys. Rev. Lett.}
  \textbf{\bibinfo{volume}{108}}, \bibinfo{pages}{156402}
  (\bibinfo{year}{2012}).

\bibitem[{\citenamefont{Kainaris et~al.}(2014)\citenamefont{Kainaris, Gornyi,
  Carr, and Mirlin}}]{PhysRevB.90.075118}
\bibinfo{author}{\bibfnamefont{N.}~\bibnamefont{Kainaris}},
  \bibinfo{author}{\bibfnamefont{I.~V.} \bibnamefont{Gornyi}},
  \bibinfo{author}{\bibfnamefont{S.~T.} \bibnamefont{Carr}}, \bibnamefont{and}
  \bibinfo{author}{\bibfnamefont{A.~D.} \bibnamefont{Mirlin}},
  \bibinfo{journal}{Phys. Rev. B} \textbf{\bibinfo{volume}{90}},
  \bibinfo{pages}{075118} (\bibinfo{year}{2014}).

\bibitem[{\citenamefont{Strunk et~al.}(2006)\citenamefont{Strunk, Stojetz, and
  Roche}}]{strunk2006quantum}
\bibinfo{author}{\bibfnamefont{C.}~\bibnamefont{Strunk}},
  \bibinfo{author}{\bibfnamefont{B.}~\bibnamefont{Stojetz}}, \bibnamefont{and}
  \bibinfo{author}{\bibfnamefont{S.}~\bibnamefont{Roche}},
  \bibinfo{journal}{Semiconductor science and technology}
  \textbf{\bibinfo{volume}{21}}, \bibinfo{pages}{S38} (\bibinfo{year}{2006}).

\bibitem[{\citenamefont{Stojetz et~al.}(2007)\citenamefont{Stojetz, Roche,
  Miko, Triozon, Forro, and Strunk}}]{stojetz2007competition}
\bibinfo{author}{\bibfnamefont{B.}~\bibnamefont{Stojetz}},
  \bibinfo{author}{\bibfnamefont{S.}~\bibnamefont{Roche}},
  \bibinfo{author}{\bibfnamefont{C.}~\bibnamefont{Miko}},
  \bibinfo{author}{\bibfnamefont{F.}~\bibnamefont{Triozon}},
  \bibinfo{author}{\bibfnamefont{L.}~\bibnamefont{Forro}}, \bibnamefont{and}
  \bibinfo{author}{\bibfnamefont{C.}~\bibnamefont{Strunk}},
  \bibinfo{journal}{New Journal of Physics} \textbf{\bibinfo{volume}{9}},
  \bibinfo{pages}{56} (\bibinfo{year}{2007}).

\end{thebibliography}

\end{document}